\documentclass[twocolumn,notitlepage,footinbib,amsmath,amssymb,amsfonts]{revtex4-1}

\usepackage{graphicx}
\usepackage{multirow}
\usepackage{color}
\usepackage{enumerate}
\usepackage[normalem]{ulem}
\usepackage{xspace}
\usepackage[hypertexnames=false]{hyperref} 

\newcommand{\mr}[1]{\ensuremath{\mathrm{#1}}}

\newcommand{\im}{\ensuremath{\mathrm{Im}}}
\newcommand{\la}{\ensuremath{\langle}}
\newcommand{\ra}{\ensuremath{\rangle}}

\newcommand{\Eq}[1]{Eq.~\eqref{#1}}

\newcommand{\Fig}[1]{Fig.~\ref{#1}}

\newcommand{\Sec}[1]{Sec.~\ref{#1}}

\newcommand{\Ref}[1]{Ref.~\cite{#1}}

\allowdisplaybreaks 

\begin{document}

\title{Filling-driven Mott transition in $\mr{SU}(N)$ Hubbard models}
\author{Seung-Sup B. Lee}
\author{Jan von Delft}
\author{Andreas Weichselbaum}
\affiliation{Physics Department, Arnold Sommerfeld Center for Theoretical Physics and Center for NanoScience, Ludwig-Maximilians-Universit\"{a}t M\"{u}nchen, Theresienstra{\ss}e 37, 80333 M\"{u}nchen, Germany}
\date{\today}
\begin{abstract}
We study the filling-driven Mott transition involving the metallic and paramagnetic insulating phases
in $\mr{SU}(N)$ Fermi-Hubbard models,
using dynamical mean-field theory (DMFT) and the numerical renormalization group (NRG) as impurity solver.
The compressibility shows a striking temperature dependence:
near the critical temperature, it is strongly enhanced in the metallic phase close to the insulating phase.
We demonstrate that this compressibility enhancement
is associated with the thermal suppression of the
quasiparticle peak in the local spectral functions.
We also explain that the asymmetric shape of the quasiparticle peak originates from the asymmetry in the underlying doublon-holon dynamics.
\end{abstract}

\maketitle
 
\section{Introduction}

The Mott insulator transition~\cite{Imada1998}, as an ubiquitous phenomenon in strongly correlated systems,
appears at the boundary of various phases.
One important example is the transition between metal and paramagnetic insulator in the Fermi-Hubbard model,
where no symmetry (e.g., spin, discrete translational invariance) is broken across the transition.
Indeed, this paramagnetic Mott transition originates
purely from the competition between Coulomb interaction
and kinetic energy where Mott insulators allow only integer number of particles
at each site, without charge fluctuations,
away from completely empty or full occupation.

Consequently, the paramagnetic Mott transition can be induced
by changing the chemical potential, i.e.,
the filling~\cite{Fisher1995,Kajueter1996,Rozenberg1997,Kotliar2002,Werner2007,Gorelik2009}.
This filling-driven Mott transition has been studied in
realizations of the Hubbard models using ultracold atoms~\cite{Jordens2008,Schneider2008,Taie2012,Duarte2015,Hofrichter2016},
in which a harmonic confinement potential is applied to the optical lattice to impose a lattice boundary.
As the confinement potential is not uniform,
the filling of each site varies from site to site
and so different phases can appear in different regions within a single trap of atoms.

However, the filling-driven paramagnetic Mott transition has not been observed yet in ultracold atom experiments,
since this requires sufficiently low temperatures
relative to the Fermi energy.
While the paramagnetic Mott insulator has been observed at temperature of about 20\% of the Fermi energy~\cite{Jordens2008,Schneider2008,Taie2012,Duarte2015,Hofrichter2016},
its evolution towards a metal occurs not via a transition but crossover at such elevated temperature.
An actual transition occurs only below the critical temperature which is only a few percent of the Fermi energy~\cite{Bluemer2002,Bulla2001,Yanatori2016}.

Recently, it was demonstrated that the one-band Hubbard model with flat potential profile can be cooled down to host the anti-ferromagnetic phase~\cite{Mazurenko2017}.
Thus one may expect that the Hubbard model with non-uniform potential also can be brought below the critical temperature of the paramagnetic Mott transition.
If so, how can one discriminate between the transition and the crossover by using quantities accessible in experiments
where temperature is not even directly measurable?

In this work, we study compressibility as a function of the occupation
number, which has been measured in ultracold atom
experiments~\cite{Duarte2015,Hofrichter2016}, in the $\mr{SU}(N)$
Hubbard model ($2 \leq N \leq 5$) with strong Coulomb interaction.  We
demonstrate that it exhibits distinct behaviors depending on whether
the temperature $T$ is below, above, or near the critical
temperature $T_c$.  To study the paramagnetic phases of the
multi-flavor model for arbitrary interaction strength, filling, and
temperature, we use dynamical mean-field theory
(DMFT)~\cite{Georges1996,Kotliar2006} and the numerical
renormalization group (NRG)~\cite{Wilson1975,Bulla2008} as its
impurity solver.

We summarize our two main results.
First, the compressibility is clearly enhanced in the
metallic phase close to the insulating phase,
i.e., when the occupation number is slightly away from integer, 
near the critical temperature.
In \Ref{Kotliar2002}, such a compressibility enhancement
(denoted as divergence therein) was first predicted
for the one- and two-band Hubbard models (i.e., $N = 2$ and $N = 4$, respectively) near half filling and explained in terms of the Landau functional.
In \Ref{Gorelik2009}, the compressibility enhancement is observed for the $N = 3$ case.
Here we generalize the scenario and provide 
a direct connection with spectral properties:
a compressibility enhancement occurs near any integer occupation
(except for completely empty and full occupation)
for flavor-symmetric Hubbard models for general $N$,
As temperature grows from $0$ to $T_c$,
the compressibility gets enhanced,
while at the same time the local spectral function gets suppressed by finite temperature.

Second, the quasiparticle peak in the metallic phase
close to the Mott transition is necessarily always strongly asymmetric:
the peak widths on positive and negative energy sides are different.
This asymmetry arises from the strong imbalance of doublon and holon occupations in the slightly doped regime.
We substantiate this argument by studying the doublon and holon correlation functions
as done in our previous studies~\cite{Lee2017,Lee2017a} on the subpeaks at the inner edges of the Hubbard bands.

The rest of this paper is organized as follows.
In \Sec{sec:method}, we provide details on the Hamiltonian and the numerical methods.
In \Sec{sec:result}, we study the compressibility, the local spectral function, and the local correlation functions of doublons and holons.
\Sec{sec:conclusion} offers our conclusion.

\section{Method}
\label{sec:method}

\subsection{$\mr{SU}(N)$ Hubbard model}
\label{sec:setup}

The $\mr{SU}(N)$ Hubbard model is the Hubbard model with $N$ flavors of fermions which are fully symmetric,
that is, Coulomb interaction strength $U$, hopping amplitude $t$, and chemical potential
are independent of particle flavor $\nu$.
Its Hamiltonian is given by
\begin{equation}
H = \sum_i \bigl[ \tfrac{U}{2} \bigl( \hat{n}_i - \tfrac{N}{2} \bigr)^2 - \mu \hat{n}_i \bigr]
- t \sum_{\la i, j\ra, \nu} ( c_{i\nu}^\dagger c_{j\nu} + \text{h.c.} ),
\label{eq:H}
\end{equation}
where $c_{i\nu}$ annihilates a particle
with flavor $\nu = 1, \ldots, N$ at lattice site $i$,
$\hat{n}_i \equiv \sum_\nu c_{i\nu}^\dagger c_{i\nu}$ is particle number operator,
$\la i,j \ra$ means nearest-neighbour sites,
and $\mu$ is the chemical potential with offset such that $\mu = 0$ yields particle-hole symmetry.
The Hamiltonian has $\mr{U}(1)$ charge and $\mr{SU}(N)$ flavor symmetry.

The $\mr{SU}(N)$ Hubbard models had been originally considered as effective descriptions of multi-band strongly correlated materials,
where the $\mr{SU}(N)$ flavor symmetry is an approximation.
Recently, these models were realized in ultracold atom experiments~\cite{Taie2012,Hofrichter2016},
where the $\mr{SU}(N)$ flavor symmetry is exact
and $N$ is tunable over the range $2 \leq N \leq 6$.

We take the chemical potential $\mu$ in \Eq{eq:H} to be uniform throughout the lattice.
Results for uniform systems are useful also for studying inhomogeneous systems within the context of the local density approximation (LDA).
A detailed analysis~\cite{Helmes2008} shows that the LDA is a good approximation
in studying both the occupation number profile (real-space distribution) and the time-of-flight (momentum-space distribution)
for the Hubbard model in a harmonic trap.

\subsection{DMFT}

Dynamical mean-field theory (DMFT)~\cite{Georges1996,Kotliar2006}
has been successfully used to study the paramagnetic Mott transition.
In the single-site setting of DMFT,
the Hubbard model is mapped onto the single-impurity Anderson model (SIAM).
There the impurity,
representing a lattice site with Coulomb interaction,
is coupled to a bath of non-interacting fermions,
and the energy dependence of the impurity-bath hybridization function encodes correlation effects (e.g., a Mott gap) within the rest of the lattice.
The self-consistent solutions of the SIAM describe {\it homogeneous} phases of the original lattice model.

The mapping onto the effective impurity model relies on the approximation 
that the self-energy is local, i.e., momentum-independent,
and charge or magnetic ordering is suppressed by assuming a fully frustrated lattice.
This approximation of locality becomes exact in the limit of infinite coordination number of lattice $z \to \infty$~\cite{Metzner1989}.
To have finite bandwidth in this limit, the hopping amplitude is scaled as $t \propto 1/\sqrt{z}$.
Then the Green's function in the lattice is derived from the impurity Green's function,
using the density of states $\rho_0 (\omega)$ of non-interacting lattice.
In this work, we consider the semi-elliptic choice $\rho_0 (\omega) = \tfrac{2}{\pi D^2} \sqrt{D^2 - \omega^2}$,
where $D \equiv 2t\sqrt{z} := 1$ is the half-bandwidth which we set as the unit of energy
Note that the Fermi energy of the non-interacting system is $D + \mu$.
We also set $\hbar = k_B = 1$ throughout.

Due to this mapping, the overall feasibility as well as the accessible parameter range of DMFT 
depend on which method is used as impurity solver to solve the effective impurity model.
Here we use the NRG as impurity solver, 
since it can provide the correlation functions on the real-frequency axis directly,
thus avoiding the numerically ill-posed problem of having to analytically continue imaginary-frequency data to the real axis.
Also, NRG is applicable to arbitrary temperature, including zero temperature 
at comparable computational cost. See the next subsection for details of the NRG method.

\subsection{NRG}

We solve the effective SIAM by using the full-density-matrix
NRG~\cite{Weichselbaum2007,Weichselbaum2012:mps}.  The bath is
discretized on a logarithmic energy grid set by the coarse-graining
parameter $\Lambda = 4$.  The resulting discrete impurity model is
mapped exactly then onto a Wilson chain with exponentially decaying
hopping.  By using energy scale separation, the iterative
diagonalization yields a complete basis of approximate many-body
eigenstates~\cite{Anders2005,Anders2006}.  Here we keep up to
$N_\mr{keep} = 2500$ low-energy multiplets at each of the early
  iterations corresponding to large energy scales.  In later
iterative diagonalization steps in the strong-coupling fixed point
regime, for computational efficiency, we also apply a rescaled
  truncation energy threshold of $E_\mr{trunc} = 9$, which is
  expected to give converged results with keeping less
  multiplets than $N_\mr{keep}$~\cite{Weichselbaum2012:mps}.  Using a
complete basis of energy eigenstates, the correlation functions at the
impurity are obtained in the Lehmann representation as a collection of
discrete spectral weights.  To recover continuous spectral functions,
we broaden the discrete spectral data with appropriate broadening
kernels~\cite{Weichselbaum2007,Lee2016}.

To simulate the multi-flavor SIAM with feasible
computational cost, we exploit the $U(1)_\text{charge} \otimes \mr{SU}(N)_\text{flavor}$ symmetry in the system
by making use of the \texttt{QSpace} tensor library
for general non-Abelian symmetries~\cite{Weichselbaum2012:sym,Weichselbaum2012:mps}.
This organizes the Hilbert space in terms of $\mr{SU}(N)$ multiplets,
and operates systematically at the level of reduced matrix elements,
with the Clebsch-Gordan coefficients split off and dealt with separately.
It allows us to efficiently perform DMFT+NRG
calculations on multi-flavor models
with $\mr{SU}(N)$ symmetry up to $N=5$,
bearing in mind that typical multiplet sizes
grow exponentially in $N$~\cite{Weichselbaum2012:sym}.
Furthermore, we use the adaptive broadening
scheme~\cite{Lee2016} to improve the spectral
resolution of correlation functions at higher energy.
Specifically, we average the results over
two discretization grids ($n_z = 2$),
followed by an adaptive log-Gaussian broadening
whose width $\sigma$ is controlled by the overall
prefactor $\alpha = 2$
and a lower bound $\sigma \geq (\ln \Lambda)/8$.
At or below the energy scale of temperature $T$ a linear broadening
is further applied to smooth out artifacts at $|\omega| \lesssim T$.
See \Ref{Lee2016} and the Supplementary Material of \Ref{Lee2017} for details.

Since the NRG calculation requires less computational cost for
  larger $\Lambda$, here we choose a rather large value $\Lambda = 4$
  to explore $O(10^4)$ data points of $(N, U, T, \mu)$ efficiently.
As a tradeoff, we have limited spectral resolution at finite energies
in the local correlation functions which is only partly regained by
using the so-called $z$-averaging procedure, standard for NRG
  applications \cite{Zitko2009,Bulla2008}, with $n_z = 2$.  In
  particular, with $\Lambda = 4$ the discretization is too crude,
e.g., to resolve the subpeaks that are known to occur at the inner
Hubbard band edges~\cite{Lee2017,Lee2017a}.  However, such features
are irrelevant for this work: The occupation number
$n = N \int_{-\infty}^\infty \mr{d}\omega \, A(\omega) / (e^{\omega/T}
+ 1)$~\footnote{%
  In the NRG, the convolution relation
  $n = N \int_{-\infty}^\infty \mr{d}\omega$
  $A(\omega) / (e^{\omega/T} + 1)$ holds when the local spectral
  function $A (\omega)$ is the discrete data in the Lehmann
  representation before broadening, not the continuous curve as
  in \Fig{fig:A}.  Since the linear
  broadening~\cite{Weichselbaum2007,Lee2016} smooths out $A(\omega)$
  for $|\omega| \lesssim T$, using the broadened $A(\omega)$ can
  introduce an artifact to the convolution relation.%
} is insensitive to sharp high-energy features in the local spectral
function
$A (\omega) \equiv A_{cc^\dagger} (\omega) = -\tfrac{1}{\pi} \im \la
c_{i\nu} || c_{i\nu}^\dagger \ra_\omega$.
And the physical phases are rather determined by the low-energy
part of $A(\omega)$, e.g., the quasiparticle peak or the Mott gap.
Moreover, we expect that the doublon-holon
subpeaks~\cite{Lee2017,Lee2017a} are fairly suppressed in the vicinity
of the Mott transition because of the strong asymmetry in the
doublon-holon dynamics to be discussed below.

In this work, we study the occupation number
$n \equiv \la \hat{n}_i \ra$,
the compressibility $\tilde{\kappa}$ [cf.~\Eq{eq:kappa} below]
and local correlation functions
$A_{XY} (\omega) \equiv -\tfrac{1}{\pi} \im \la X_{i\nu} || Y_{i\nu} \ra_\omega$,
which are the imaginary part of retarded time correlators of local operators $X$ and $Y$ acting on site $i$, transformed to the frequency domain.
Based on the DMFT mapping onto the SIAM and the semi-elliptic density of states $\rho_0 (\omega)$,
these local properties at a lattice site are equivalent to the same properties at the impurity
when the self-consistent solution of the SIAM is achieved.

\subsection{Compressibility}

\begin{figure}
\centerline{\includegraphics[width=.49\textwidth]{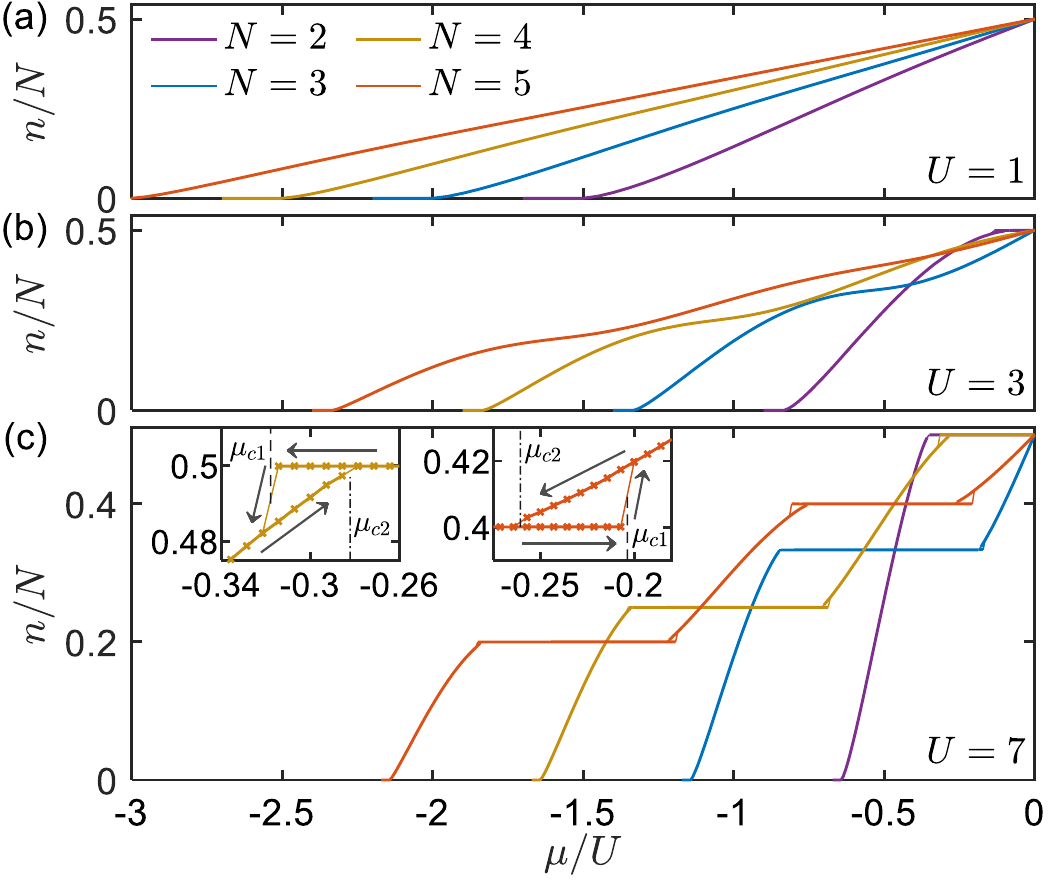}}
\caption{
Particle number per site $n \equiv \la \hat{n}_i \ra$ vs.~chemical potential $\mu$
along the homogeneous, paramagnetic phases of the $\mr{SU}(N)$ Hubbard models [cf.~\Eq{eq:H}] at temperature $T = 0$ (thick lines).
Due to particle-hole symmetry at $\mu=0$,
the curves for $\mu > 0$ can be deduced by $n (\mu) = N - n (-\mu)$.
(a) For small $U$, the systems remain compressible, i.e., metallic.
(b) For intermediate $U = 3$, plateaus start to develop at integer $n$.
(c) For large $U = 7$, wide plateaus
demonstrate the incompressibility of the Mott insulating phase.
As the flat plateaus for the insulating phase connect to the slanting lines for the metallic phase near the Mott transition,
weak hysteretic behavior occurs which thus indicates coexistence.
Insets: Zoom-in to individual
hysteresis loops for $N = 4$ and $5$,
as examples for the left and right ends of the Mott plateaus, respectively.
Each thin solid line connects two data points
(crosses) across the Mott transition:
For the insulator-to-metal transition (IMT) at $\mu_{c1}$ (dashed vertical lines), it connects
the last data point in a plateau, with the subsequent next
point in the metallic phase. Conversely, for the
metal-to-insulator transition (MIT) at $\mu_{c2}$ (dash-dotted vertical lines) it connects the last point
in the metallic phase (slanted line) with the next
data point in the insulating plateau (here at $T=0$,
these are very short line segments visible only in the insets).
}
\label{fig:T0_n}
\end{figure}

The compressibility is defined as 
\begin{equation}
	\tilde{\kappa} \equiv \frac{\partial n}{\partial \mu} \equiv n^2 \kappa \text{,}
	\label{eq:kappa}
\end{equation}
where we only use the derivative $\tilde{\kappa}$, without rescaling,
for the remainder of the paper \cite{Hofrichter2016}.
We obtain $n$ for a linear grid of $\mu$ with grid size $\Delta \mu = 0.05$,
and compute $\tilde{\kappa}$ by numerically
differentiating $n (\mu)$.
Since the latter is sensitive to numerical noise,
even if the curves $n (\mu)$ look smooth except at phase transition points (cf.~\Fig{fig:T0_n}),
we determine the slope of $n(\mu)$ at $\mu = \mu'$
by fitting at most five consecutive points on the short interval $\mu' - 2\Delta \mu \leq \mu \leq \mu' + 2\Delta \mu$
with a quadratic polynomial.
When $U$ is larger than the critical strength,
a Mott transition occurs, signaled by discontinuities in $n (\mu)$ and/or $\tilde{\kappa} (n)$,
as discussed in much detail in the following sections.
Thus, when $\mu'$ is close to these values,
we exclude the points $n (\mu)$ beyond the critical value
to keep the fitting error minimal.

\begin{figure}
\centerline{\includegraphics[width=.49\textwidth]{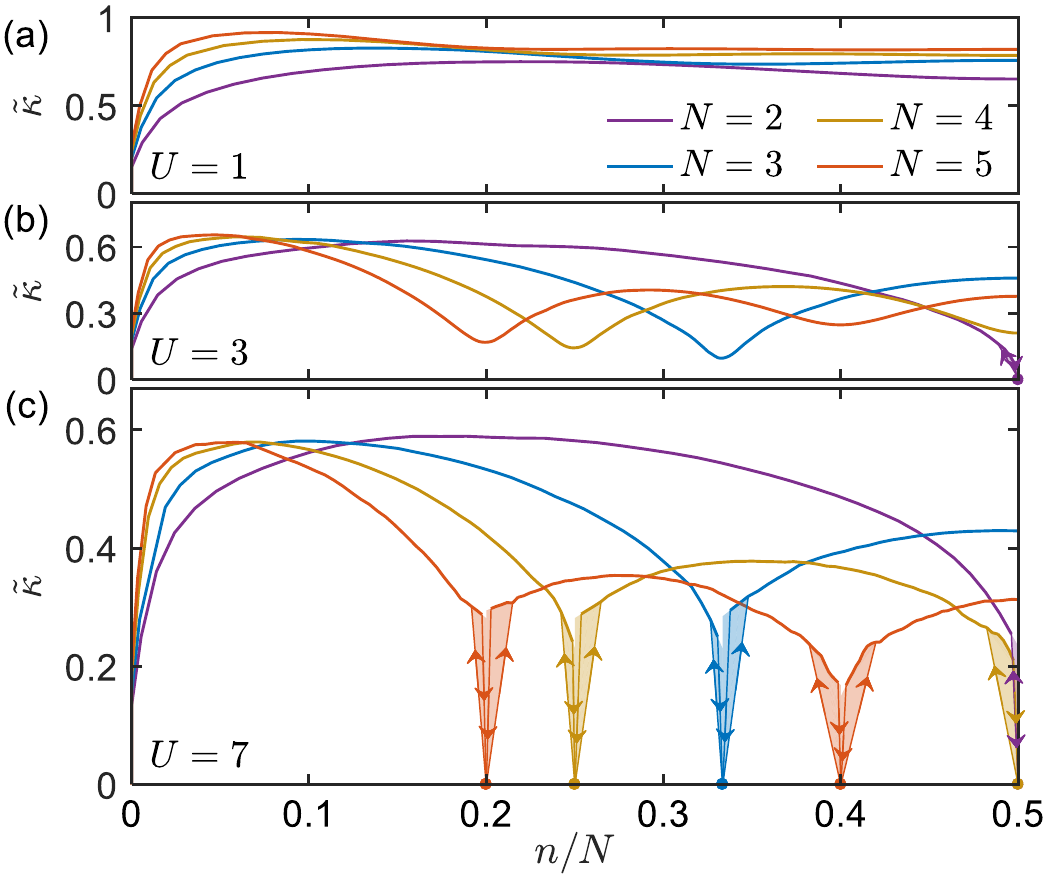}}
\caption{ Compressibility $\tilde{\kappa}$ of Eq.~\eqref{eq:kappa}, 
obtained as numerical derivative from the
curves in \Fig{fig:T0_n} at $T=0$, as function of filling fraction
$n/N$ (thick solid lines).  In panels (b) and (c),
thin lines with arrows indicate discontinuous jumps between finite $\tilde{\kappa}$ in
    the metallic phase and $\tilde{\kappa} = 0$ in the insulating
    phases.  Thus the boundary of each shaded area represents a
    hysteresis loop, where the upper solid line part of the
    boundary indicates the metallic solution in the coexistence
    region.  Note that the shading just below $n=1$
    for $(N, U) = (2, 7)$ is not visible, since the coexistence region
    $[\mu_{c1},\mu_{c2}]$ in the curve $n(\mu)$ is narrower than the
    grid size $\Delta \mu = 0.05$.  To indicate the second-order
    nature of MIT at $T = 0$, we have extrapolated the shaded areas to
    integer $n$, while the lines with downward arrows with
finite slopes connect actual data points. The curves for $n > N/2$ can
    be deduced by particle-hole symmetry
    $\tilde{\kappa}(n) = \tilde{\kappa}(N-n)$. }
\label{fig:T0_dn}
\end{figure}

\section{Results}
\label{sec:result}

\subsection{Zero temperature}

We start with studying the filling-driven Mott transition at $T = 0$,
which by definition is free from thermal fluctuations.
These results are obtained by directly solving the SIAM at infinitesimally
low temperature $T = 0^+$, i.e., not by extrapolating finite-$T$ results~\cite{Werner2007}.
Indeed, this accessibility of low temperatures is a major strength
of using NRG as the DMFT impurity solver~\cite{Bulla1999}.

In \Fig{fig:T0_n} we present our data on the particle number per
site $n$ vs.~the full range of chemical potential $\mu$, and in
\Fig{fig:T0_dn} we depict the corresponding compressibility
$\tilde{\kappa}$ vs.~filling $n / N$.  For weak interaction $U = 1$,
the compressibility $\tilde{\kappa}$ is almost independent of $n$ for
$n / N \gtrsim 0.1$.  In contrast, $\tilde{\kappa} (n)$ has local
minima at integer $n$ for larger $U \geq 3$.  In \Fig{fig:T0_n},
  lines $n(\mu)$ with finite slope (i.e., $\tilde{\kappa} > 0$)
  correspond to metallic phases.  Conversely, horizontal plateaus
  represent the incompressible phase of a Mott insulator. These
  plateaus appear due to an interaction-driven Mott transition,
  i.e., by increasing $U$ beyond a critical interaction strength
  $U_c (N, [n])$ that depends on both the number of flavors
  $N$~\cite{Ono2003,Inaba2005,Bluemer2013} as well as the integer filling $[n]$ of the
  Mott plateau~\cite{Rozenberg1997}.

At each end of a plateau in the $n(\mu)$ curves
(except for $n = 0, N$),
a Mott transition occurs, accompanied with a hysteresis
loop~\cite{Kotliar2002,Werner2007}, as shown (in the insets) in
\Fig{fig:T0_n}(c).  For each hysteresis loop, we can define a pair of
critical values $\mu_{c1}$ and $\mu_{c2}$ of the chemical
potential: $\mu_{c1}$ is the value of the chemical potential at
the outer edge of a plateau in $n(\mu)$ which thus describes an
insulator-to-metal transition (IMT).  Similarly, $\mu_{c2}$
is the value where the metallic solution
terminates within a Mott plateau, and thus describes a
metal-to-insulator transition (MIT).  Therefore, in between two
critical values ($\mu_{c1} < \mu < \mu_{c2}$ and
$\mu_{c2} < \mu < \mu_{c1}$ for the left and right ends of the Mott
plateau, respectively), both insulating and metallic phases coexist,
i.e., $n(\mu)$ is double-valued.

The compressibility $\tilde{\kappa}$ vs. $n$ also shows
  discontinuities at integer $n$, associated with the Mott transition,
  as can be observed in Fig.~\ref{fig:T0_dn}(b) and (c).
An IMT, depicted by an upward arrow, involve not only a jump in $\tilde{\kappa}$ but also in the occupation $n$.
Similarly, also across a MIT, depicted by a downward arrow, 
a jump in both $\tilde{\kappa}$ \emph{and} $n$ can occur.
However, at $T=0$, the jump in $n$ should disappear, 
such that $n$ evolves \textit{continuously} across MIT~\cite{Werner2007}.
Thus the downward arrows in \Fig{fig:T0_dn} should in principle be strictly vertical;
the reason why they are slanted, instead, is the nonzero grid size, $\Delta \mu = 0.05$, used for our calculations.

The continuity of $n$ and the discontinuity of $\tilde{\kappa}$
  across the metal-to-insulator transition reflect the second-order
  nature of the Mott transition at $T = 0$.  Within the coexistence
  regime, there exists another critical value of chemical potential
  $\mu_c$ at which the metallic and insulating solutions have the same
  values of free energy.  For finite temperature $0 < T < T_c$, the
  transition at $\mu = \mu_c$ is first-order.
In contrast, for $T = 0$, one has $\mu_c = \mu_{c2} \neq \mu_{c1}$ and the transition
  at $\mu = \mu_c$ turns into a second-order transition~\cite{Kotliar2002,Werner2007}, 
in that $n$ is continuous but $\tilde{\kappa}$ is discontinuous 
(note that $n$ and
  $\tilde{\kappa}$ are proportional to the first and second
  derivatives of the free energy with respect to chemical
  potential~\cite{Werner2007}).  Another exceptional 
  situation for which the transition at $\mu = \mu_c$ is not first-order
  arises at the critical end point $T = T_c$,
  where the coexistence region shrinks to a point, i.e.,
  $\mu_c = \mu_{c1} = \mu_{c2}$, and, again, the transition
  becomes second-order.

For $n = 0$ and $n = N$ the system is a band insulator,
and thus no longer a Mott insulator. Correspondingly,
we also observe no phase coexistence near the plateaus for $n = 0$ and $N$.
Therefore while these trivial phases are still incompressible,
their plateaus are excluded from our discussion of Mott plateaus.
The value for the chemical potential below which
the system becomes empty, is given by $\mu \leq
\mu_0 \equiv -(N-1)\tfrac{U}{2} - D$,
in agreement with the overall trend seen in \Fig{fig:T0_n}.
This value can be motivated as follows:
By substituting $\mu' = -(N-1)\tfrac{U}{2}$ to $\mu$,
the first term in \Eq{eq:H}
favors zero and one occupation numbers equally.
By adding another shift in the
chemical potential, $\mu''=-D$, when associated with the
second term in \Eq{eq:H}, this empties
this non-interacting kinetic part of the Hamiltonian.
Therefore the system becomes empty ($n=0$)
for $\mu \leq \mu_0 \equiv \mu'+\mu''$, resulting in the
above expression.
Similarly, by particle-hole transformation, the system becomes completely filled ($n=N$) for
$\mu \geq -\mu_0$.

\begin{figure}
\centerline{\includegraphics[width=.49\textwidth]{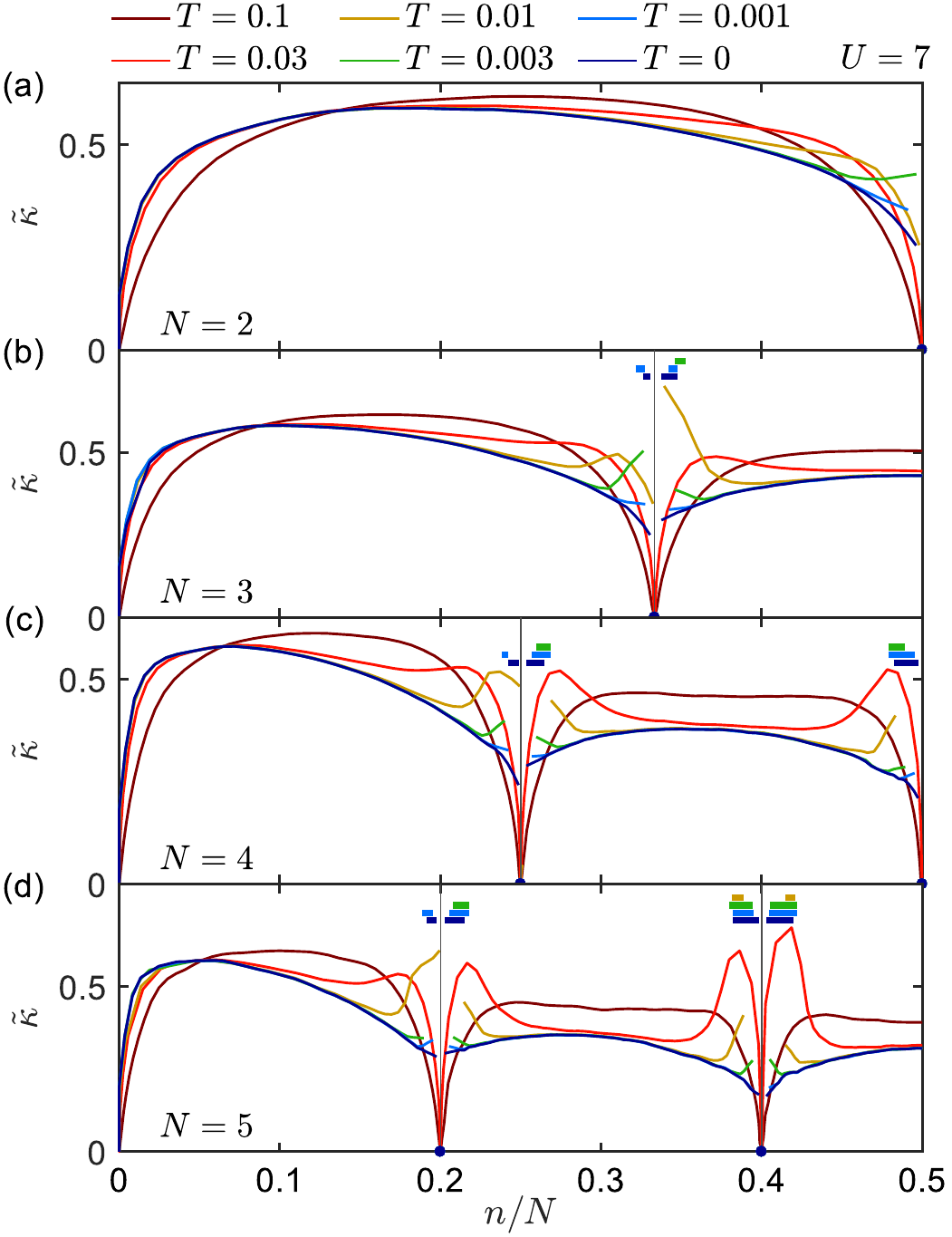}}
\caption{
Compressibility $\tilde{\kappa}$ vs.~filling fraction $n/N$
for fixed $U = 7$ and varying $T$, with integer fillings
indicated by vertical lines.
Below the critical temperature $T_c (N,U,\mu) \in (0.01, 0.03)$,
there are Mott transitions appearing as discontinuous
jumps in $\tilde{\kappa}$.
Thick horizontal bars above the curves $\tilde{\kappa} (n)$ (solid lines)
near integer $n$ (vertical gray lines)
indicate the range of $n$ for the metallic phase in the coexistence regime,
color matched with the curves $\tilde{\kappa}$ for the same $T$.
As $T$ increases from $0$ to $T_c$,
the coexistence regions get narrower in $n$.
For some curves, e.g., for $N = 2$, the coexistence region
is narrower than the numerical grid size $\Delta \mu = 0.05$.
As $T$ is increased towards $T_c$,
local maxima of $\tilde{\kappa} (n)$ appear near integer $n$,
which we call compressibility enhancement.
At $T > T_c$, the discontinuity in the curves $\tilde{\kappa} (n)$ disappears,
which indicates that a crossover (rather than a
phase transition) occurs between metallic and insulating behavior.
}
\label{fig:Tdep_dn}
\end{figure}

\begin{figure}
\centerline{\includegraphics[width=.49\textwidth]{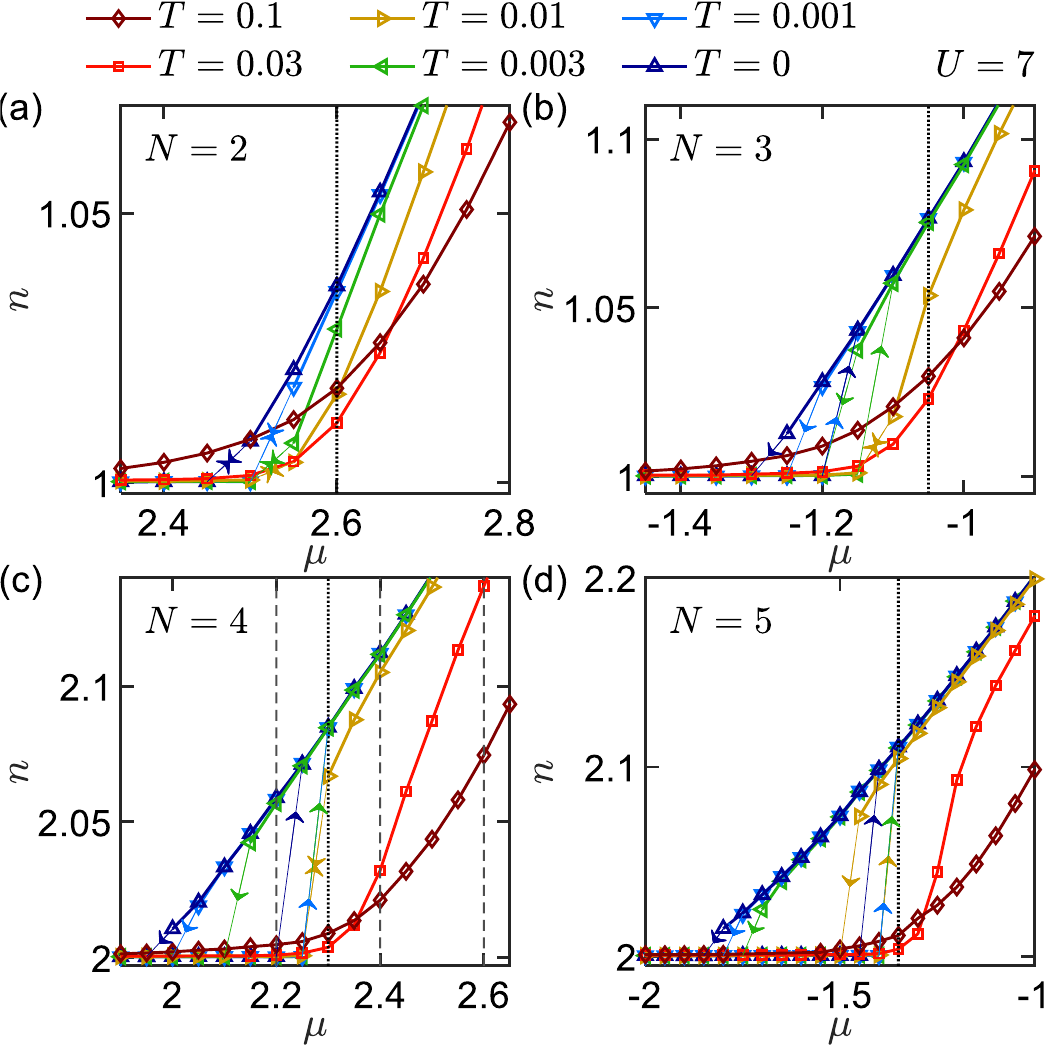}}
\caption{
Particle number per site $n$ vs.~chemical potential $\mu$ zooming into the Mott transitions
just above $n = \lfloor N/2 \rfloor$,
i.e., particle-doped regime,
for $U = 7$ and $2 \leq N \leq 5$.
Thick solid lines and symbols represent $n (\mu)$.
Thin lines connect data points in different
phases, where the arrow
specifies the direction of phase transition.
For the values of $\mu$ indicated by vertical dotted lines,
we show the local spectral functions in \Fig{fig:A} below.
In \Fig{fig:A2}, 
we illustrate the local spectral functions for $N = 4$,
with the $\mu$ values marked by vertical dashed lines in panel (c).}
\label{fig:Tdep_n}
\end{figure}

\subsection{Finite $T$ and compressibility enhancement}
\label{sec:finT}

Next we analyze the effect of finite temperature $T$ on the
  compressibility curves $\tilde{\kappa} (n)$, as shown in
\Fig{fig:Tdep_dn}.  We observe that the jumps in
$\tilde{\kappa} (n)$ near integer $n$ survive for temperatures below a
critical value $T_c$, and disappear above it.  That is, $T_c$ is the
critical temperature of the paramagnetic Mott transition, and it
depends on $N$, $U$, and $\mu$.  For the $U = 7$ considered in
\Fig{fig:Tdep_dn}, we find that $T_c$ lies between 0.01 and 0.03.

Near and across the critical temperature, $T \sim T_c$, we
observe a \textit{compressibility enhancement}: $\tilde{\kappa} (n)$
exhibits local maxima for $n$ close,  but not equal to, integer values.
These local maxima of $\tilde{\kappa}(n)$ become more pronounced as
$T$ gets closer to $T_c$,  both from above and below.  For
example, for the curves of $N=5$ and $T\in [0.01,0.03]$, a peak of
$\tilde{\kappa}(n)$ associated with the compressibility enhancement is
even the global maximum, not only a local maximum.  In contrast, for
$T = 0$ and $T=0.1$, which are far below and above $T_c$,
respectively, the curves $\tilde{\kappa} (n)$ decrease monotonically as
$n$ approaches an integer both from above or below, and reach
  zero either by a jump for $T < T_c$ or continuously for $T > T_c$.

The compressibility enhancement directly originates, by definition,
from qualitative changes in the curves $n(\mu)$ for different $T$.
In \Fig{fig:Tdep_n}, we plot $n (\mu)$ for the same choice of
parameter sets $(N, U, T)$ as in \Fig{fig:Tdep_dn},
but zooming in towards the
coexistence region of the Mott transitions,
choosing $n$ slightly larger than $\lfloor N/2 \rfloor$,
i.e., the down-rounded value of $N/2$.
As $T$ increases from $0$ to $T_c$,
the inner edge of the coexistence region, $\mu_{c2}$,
rapidly shifts towards the outer edge, $\mu_{c1}$,
while $\mu_{c1}$ likewise shifts outward, but slower than $\mu_{c2}$ does.
Hence the width of the coexistence regime, $| \mu_{c1} - \mu_{c2} |$, decreases.
In a sense, therefore, finite temperature destabilizes the metallicity,
but stabilizes the insulting phase.
Interestingly, this behavior
is similar to the $T$-dependence
of $U_{c2}$ and $U_{c1}$ at half filling~\cite{Bulla2001,Bluemer2013}.
At the same time,
the slope of the slanted part of $n(\mu)$ (corresponding
to the metallic phase) in the close vicinity to integer $n$ increases,
leading to the compressibility enhancement in \Fig{fig:Tdep_dn}.
We also see that  contrary to the $T = 0$ case,
not only $\tilde{\kappa}$ but also $n$ exhibits
a jump at $\mu = \mu_{c2}$ for $0 < T < T_c$.

Finally let us emphasize that while the data in \Fig{fig:Tdep_n}
focusses on the particle-doped case (with $n$ slightly
above $\lfloor N/2 \rfloor$), the overall behavior for
the hole-doped case ($n$ slightly below $\lfloor N/2 \rfloor$) is completely analogous, bearing in mind that,
via particle-hole transformation, one has $n(\mu) = N - n(-\mu)$.

\begin{figure}
\centerline{\includegraphics[width=.49\textwidth]{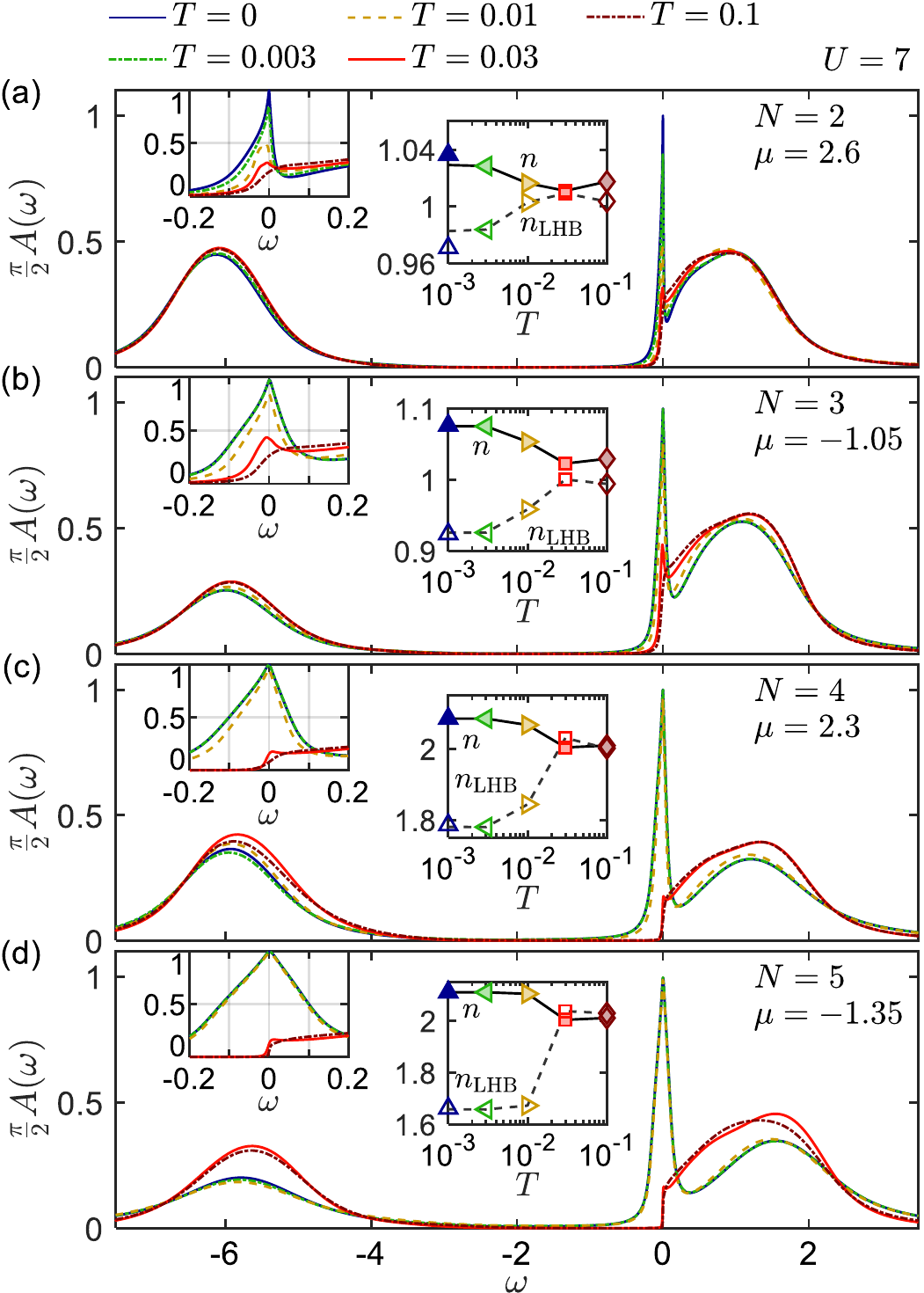}}
\caption{
Local spectral functions $A(\omega)$ in the metallic phase
with slight particle doping, for large interaction $U = 7$,
with varying parameters $(N, T, \mu)$, where $\mu$ corresponds to the
values marked by the vertical dotted lines in \Fig{fig:Tdep_n}.
Left insets zoom into the low-frequency regime containing the quasiparticle peak.
Middle insets show how the occupation number $n$ (filled symbols)
together with the spectral weight of the lower Hubbard band, 
$n_\mr{LHB} \equiv N \int_\mr{LHB} \mr{d}\omega \, A(\omega)$ (empty symbols), change with $T$,
where for the large value of $U$ here we delineate the range of 
the LHB by $\omega<-2$.
For reference, the dark blue symbols on top of the left axis in the middle insets
give $n$ and $n_\mr{LHB}$ for $T = 0$, and not for $T = 10^{-3}$.
}
\label{fig:A}
\end{figure}

\subsection{Local spectral functions}

The coexistence region analyzed in \Fig{fig:Tdep_n} is intricately
linked to a competition between the metallicity that permits
non-integer occupation, and the Mottness that constrains $n$ to be
integer. Therefore in order to gain a better understanding, we
now look into the local spectral functions in the metallic
regime, with a focus on the mutual interplay between average local
occupation $n$ and the quasiparticle peak in the spectral data.

The decrease of the average local occupation $n$ towards the
Mott plateau $[n]$ (the rounded value of $n$) as $T$ is
increased, as shown in \Fig{fig:Tdep_n}, is necessarily
connected to the thermal suppression of the quasiparticle peak in the
local spectral function $A(\omega)$. Since its total weight,
$\int_{-\infty}^\infty \mr{d}\omega \, A(\omega)=1$, is
preserved by a sum rule, the ensuing transfer of spectral
weight necessarily also influences the local occupation,
$n = N \int_{-\infty}^\infty \mr{d}\omega \, A(\omega) / (e^{\omega/T}
+ 1)$.

A detailed analysis of the spectral data in the metallic
phase is presented in \Fig{fig:A}.  There we show in each panel
the local spectral functions for fixed $N$ and $\mu$ but
for several values of $T$. Because of the large $U = 7$,
there are two well-developed Hubbard bands, 
centered at $\omega \sim D - U$ and $\omega \sim D$,
respectively.  Since $\mu$ is chosen
as a fixed value slightly larger than the critical values $\mu_{c2}$
for different $T$, the spectral data is strongly asymmetric
around $\omega=0$ despite, e.g., $n \approx N/2$ for even
$N$.  Specifically, the lower Hubbard band is well separated
towards negative frequencies, whereas the lower edge of the upper
Hubbard band is close to the Fermi level.

In addition to the Hubbard bands, the spectral functions in
\Fig{fig:A} for $T \leq 0.01 < T_c$ feature a quasiparticle peak at
the Fermi level $\omega = 0$.  As $T$ increases, the quasiparticle
peak gets suppressed and the occupation number $n$ approaches
$[n]$ (see middle insets).  The quasiparticle peaks for
$T < T_c$ represent Fermi-liquid quasiparticles which, due to the
narrowness of the peak, have heavy effective mass.  This Fermi-liquid
state hosts low-energy charge fluctuations, so non-integer
$n$ is generally possible~\cite{Lee2017}.

On the other hand, for $T \geq 0.03 > T_c$, the significant
suppression or the absence of the quasiparticle peak rules out
coherent low-energy quasiparticles.  So the state of the system is
well described by the Hubbard bands only. If the lower
Hubbard band (LHB) below the Fermi level is fully occupied and the
upper one (UHB) above the Fermi level is empty, the lattice sites
are filled by an integer number of particles, $[n]$, without
charge fluctuations.  Accordingly, the integrated spectral
weights of the individual Hubbard bands would be also integers,
e.g., $n_\mr{LHB} \equiv N \int_\mr{LHB} \mr{d}\omega \, A(\omega) = [n]$.
In \Fig{fig:A}, however where $\mu$ has been chosen
slightly above $\mu_{c1}$, the lower edge of the upper
Hubbard band has dropped slightly below the Fermi level,
thus making a small contribution to the occupancy. As
a consequence, the values $n (T = 0.03)$ are very close
to, but slightly larger than, integers $[n]$.  For
even larger temperature $T = 0.1$, the thermal window of the
Fermi-Dirac distribution $(e^{\omega/T} + 1)^{-1}$ widens, so the
  occupation $n$ deviates even more strongly from the integer $[n]$.

The $T$-dependence of $n$ and $n_\mr{LHB}$, presented in the
middle insets of \Fig{fig:A}, show how the quasiparticle
weight, i.e., the spectral weight of the quasiparticle peak, is
transferred to the Hubbard bands as $T$ increases. Since
the total spectral weight is preserved, the difference
$\bigl[n(0) - n_\mr{LHB}\bigr]_{T=0}$ is equivalent to the
negative-frequency part of the quasiparticle weight, i.e.,
integrated up to $\omega=0$.  This weight is fully transferred
to the Hubbard side bands as $T$ is increased towards $T_c$.
This spectral weight transfer can be split into two net
flows: the weight $[n] - n_\mr{LHB}\big|_{T=0}$ is transferred to the LHB,
whereas the rest, i.e., $n\big|_{T=0} - [n]$, together with the quasiparticle
weight for $\omega > 0$, flows into the UHB.  Surprisingly, despite
the distance between the LHB and the Fermi level, a significant
portion of the negative-frequency quasiparticle weight flows into the LHB.
[e.g., see the change of height in the LHB in \Fig{fig:A}(d)].
  
The thermal suppression of the quasiparticle peak, which is
accompanied by a transfer of spectral weight, thus
pushes $n$ closer towards $[n]$ with increasing $T$ at fixed $\mu$.
Correspondingly, $\mu_{c2}(T)$ changes much more
sensitively with $T$ than $\mu_{c1}(T)$.
In the metallic phase near $\mu_{c2}(T)$,
the transfer of spectral weight from the quasiparticle peak into
the LHB with increasing $T$ necessarily leads to an increase in
$\mu_{c2}$, whereas no such weight-transfer occurs near
$\mu_{c1}(T)$ in the insulating phase, which lacks a quasiparticle
peak. This is consistent with the fact, discussed in
\Sec{sec:finT},  that increasing $T$ 
causes a stronger increase in $\mu_{c2}(T)$ than in $\mu_{c1}(T)$.

\begin{figure}
\centerline{\includegraphics[width=.49\textwidth]{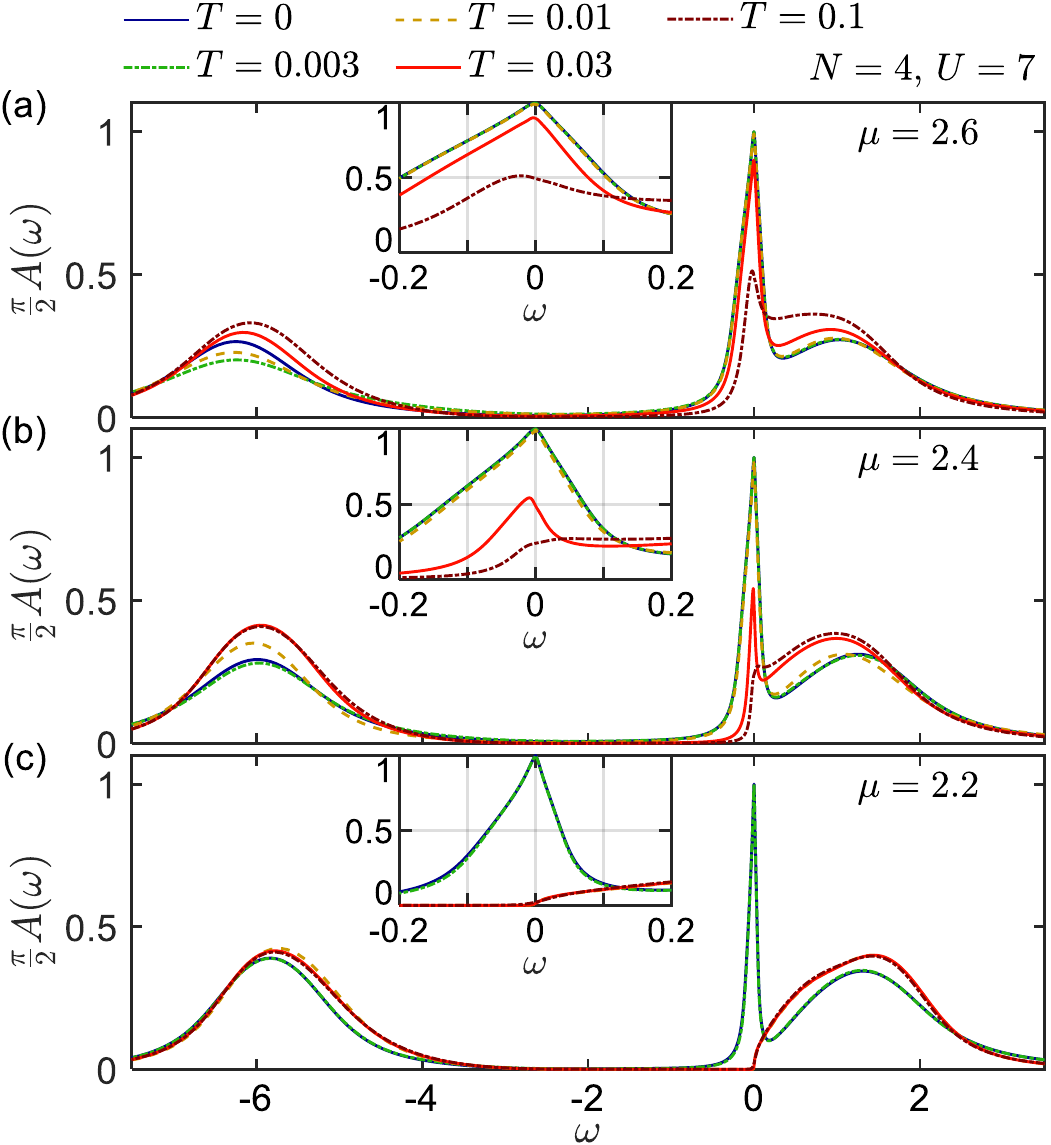}}
\caption{
Evolution of the local spectral functions $A(\omega)$ for
$N = 4$, $U = 7$, and a set of five
fixed temperatures,  as $\mu$ is decreased (top to
bottom) towards the MIT at $\mu_{c2}$.  The three  $\mu$-values 
shown here all lie in the vicinity 
of the value $\mu = 2.3$ of \Fig{fig:A}(c).
Insets zoom into low-frequency regime in
which the quasiparticle peak or the Mott gap appears.  Note that
the curve for $T = 0.01$ has no quasiparticle peak in panel (c),
since $\mu = 2.2$ lies below $\mu_{c2}(T = 0.01) = 2.275(25)$, i.e., $\mu$ has already been
lowered past the MIT transition point.
}
\label{fig:A2}
\end{figure}

Once the chemical potential is outside the range of the
Mott plateau (including the coexistence region), the system
is always metallic, and the spectral functions evolve
smoothly in terms of a crossover as temperature increases.
Therefore in this case, quasiparticle peaks
are present also for temperatures above $T_c$,
which denotes the critical temperature right at the IMT and MIT.
This is the reason why quasiparticle peaks occur
for $T = 0.03 > T_c$ in \Fig{fig:A}(a)-(b).

However, for these high-temperature peaks the evolution with
decreasing $\mu$ at fixed $T$ is qualitatively different from those
of the low-temperature peaks in the regime $T \le 0.01 < T_c$ within the metallic
phase.  
This is illustrated in \Fig{fig:A2}, which shows how
the local spectral function $A(\omega)$ evolves when $\mu$ is
decreased (top to bottom) towards the MIT at $\mu_{c2}$. As the MIT
is approached while lowering $\mu$ at a given, fixed temperature,
the quasiparticle peak behaves differently depending on whether that
temperature lies above or below $T_c$.  In the former case, i.e., if
the fixed temperature satisfies $T \geq 0.1 > T_c$, both the height
and width of the quasiparticle peak smoothly decrease with
decreasing $\mu$, which is consistent with the crossover behaviour
above $T_c$.  In contrast if the fixed temparature lies in the
range $T \leq 0.01 < T_c$, the quasiparticle peak becomes narrower
as $\mu$ decreases towards $\mu_{c2}$, while its height remains almost
unchanged.  Once $\mu$ has passed below the MIT at $\mu_{c2}$,
the height $A (\omega = 0)$ drops abruptly, which is consistent with
the transition nature below $T_c$.  In the limiting case of
$T = 0$, the spectral function $A(\omega = 0)$ is pinned
to the value $2/\pi$ all along the metallic phase, as dictated by
the Luttinger theorem~\cite{Mueller-Hartmann1989}.
Our DMFT+NRG result in \Fig{fig:A} fulfills this
relation with accuracy better than 3\%
due to the intrinsic high accuracy of NRG at low energies,
despite strongly broken particle-hole symmetry.
For example, at $T = 0$ the curves in \Fig{fig:A} 
have the zero-frequency values 
$\frac{\pi}{2} A(\omega = 0) \simeq 
0.9989$, $0.9726$, $0.9998$, $0.9925$
for $N=2$, $3$, $4$, $5$, respectively.

Again let us emphasize that, while the spectral data
in \Fig{fig:A} above is for the particle-doped case,
the spectral functions for slight hole doping can be simply deduced
by particle-hole transformation, which yields the equivalence
$A (\omega) |_\mu = A(-\omega) |_{-\mu}$.

We briefly discuss the effect of a non-uniform potential.  In
  \Ref{Helmes2008}, the paramagnetic Mott transition of the
  $\mr{SU}(2)$ Hubbard model has been studied by using real-space
  DMFT, which incorporates the non-uniformity of harmonic confinement
  potential.  There the metallic phase was found to exist in a
  wider region than predicted by the LDA (which we use in this work;
  see \Sec{sec:setup}), since the metallicity can ``penetrate'' into
  nearby insulating regions via the Kondo effect.  On the other hand, the
  relation between the deviation of the
  local occupation number from integer and the thermal suppression of
  the quasiparticle peak was also found there, 
(e.g., see Fig.~5 of \Ref{Helmes2008}),
consistent with our result.

\subsection{Doublon and holon correlators}
\label{sec:DH}

The overall spectral data at finite doping as in \Fig{fig:A}, by
construction, is always strongly particle-hole asymmetric.  At weak
doping and large $U$, the quasiparticle peak is necessarily close to
one Hubbard band but clearly separated from the other. This
asymmetry is also reflected in the shape of the quasiparticle peak
itself.

\begin{figure}
\centerline{\includegraphics[width=.49\textwidth]{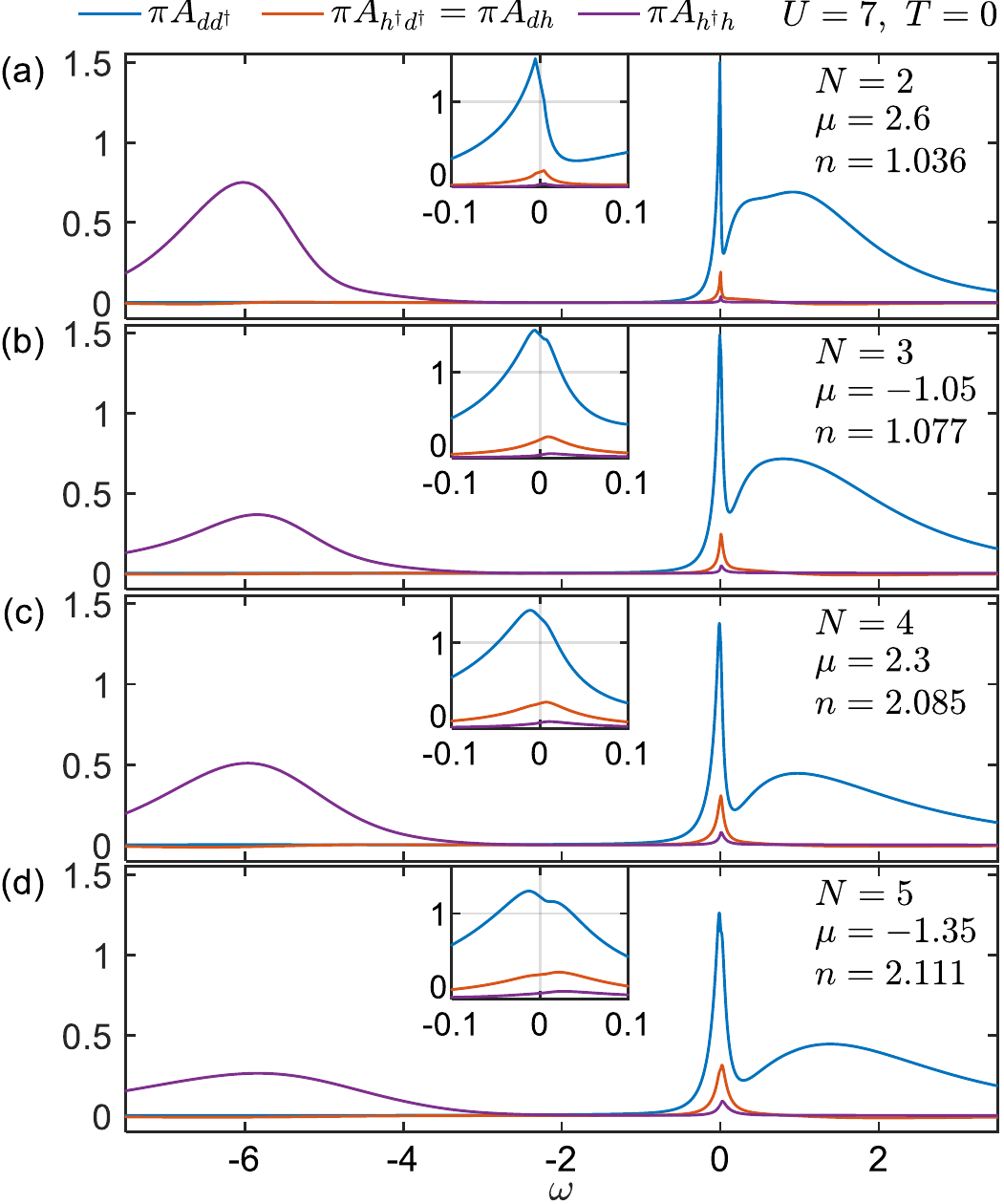}}
\caption{
The correlation functions of doublons and holons at $T = 0$
that correspond to the local spectral functions for slight
particle doping at $T = 0$ in \Fig{fig:A}.  Insets zoom into the
low frequency region containing the quasiparticle peak.
Among four correlators involving doublons and holons, the
doublon-doublon correlator $A_{dd^\dagger}$ makes the largest
contribution to the quasiparticle peak, while the other
correlators have much less spectral weight in the region of 
the quasiparticle peak.
}
\label{fig:DH}
\end{figure}

It is possible to understand the origin of this asymmetry
of the quasiparticle peaks by studying the correlation
functions of doublons and holons~\cite{Lee2017}.
We define a doublon (holon) as a local excitation that creates (annihilates) a particle at a lattice site filled by $[ n ]$ particles.
Accordingly the creation operators for doublons and holons are expressed as the projected operators (so-called Hubbard operators),
\begin{equation}
d_{i\nu}^\dagger = c_{i\nu}^\dagger P_{i,[n]}, \quad
h_{i\nu}^\dagger = c_{i\nu} P_{i,[n]},
\label{eq:dh}
\end{equation}
where $P_{i,[n]}$ is the projector onto the subspace
in which site $i$ has $[n]$ particles.
For the $N = 2$ case, the fermion operator $c_{i\nu}$ can be decomposed into doublon and holon operators,
$c_{i\nu} = d_{i\nu} + h_{i\nu}^\dagger$.
Thus the local spectral function can be decomposed as
\begin{subequations} 
\begin{equation}
A (\omega) = A_{dd^\dagger} (\omega) + A_{h^\dagger d^\dagger} (\omega) + A_{d h} (\omega) + A_{h^\dagger h} (\omega),
\label{eq:Aomega}
\end{equation}
where the relation $A_{h^\dagger d^\dagger} (\omega) = A_{d h} (\omega)$ holds generally.
At the particle-hole symmetric point $\mu = 0$, we have the further relation
$A_{dd^\dagger} (\omega) = A_{h^\dagger h} (-\omega)$.
As before, the doublon and holon correlators for slightly hole doped cases
can be deduced from the particle doping result in \Fig{fig:DH} via particle-hole symmetry:
\begin{align}
A_{dd^\dagger} (\omega) |_\mu & = A_{h^\dagger h} (-\omega) |_{-\mu} , 
\\ \nonumber
A_{h^\dagger d^\dagger} (\omega) |_\mu & = A_{dh} (\omega) |_\mu = A_{h^\dagger d^\dagger} (-\omega) |_{-\mu} = A_{dh} (-\omega) |_{-\mu}.
\end{align}
\end{subequations}
For $N > 2$, the decomposition of $c_{i\nu}$ involves other projected operators, 
such as $c_{i\nu} P_{i,[n]-1}$, in addition to $d_{i\nu}$ and $h_{i\nu}^\dagger$.
In this case, the decomposition of $A(\omega)$
in \Eq{eq:Aomega} is not exact but approximate,
since additional terms arise beyond those shown
in \Eq{eq:Aomega}.
However, these additional terms are negligible for large $U$.
Indeed, the deviation $| A (\omega) - \sum_{X,Y = d,h^\dagger} A_{XY^\dagger} (\omega) |$
becomes smaller,
since the probability that a site contains less than $[n] - 1$ or more than $[n] + 1$
particles is suppressed due to the large cost in Coulomb energy.
Therefore the projected particle operators which involve the projectors
$P_{i,m}$ with $m > [n]+1$ or $m < [n]-1$
(i.e., other than doublon and holon operators) have
negligible contribution to the correlation functions.

Before analyzing specific results,
let us discuss a few general properties of the correlators of doublons and holons.
From \Eq{eq:Aomega}, we have
three independent doublon and holon correlators,
$A_{dd^\dagger} (\omega)$, $A_{dh} (\omega) =
A_{h^\dagger d^\dagger} (\omega)$, and $A_{h^\dagger h} (\omega)$,
as shown in \Fig{fig:DH}.
They reflect the overall structure of the full
spectral function, including the Hubbard bands and the quasiparticle peak in the
metallic phase~\cite{Lee2017}.
The doublon-doublon correlator $A_{dd^\dagger}$ and
the holon-holon correlator $A_{h^\dagger h}$ have
two major features. $A_{dd^\dagger}$ ($A_{h^\dagger h}$) 
carries most of the UHB (LHB),
as well as the negative (positive)
frequency side of the quasiparticle peak, respectively.
The latter are centered around a small energy scale
$-\omega_s$ ($\omega_s$),
where $\omega_s$ corresponds to the width of the
quasiparticle peak centered around $\omega=0$,
which itself can be related to the energy
scale of flavor-like collective modes~\cite{Lee2017a}
via the (dynamical) flavor susceptibility.
Note that, for $N = 2$, flavors equivalently represent spins for one band of electrons.

The features in the doublon-holon correlation functions
necessarily correspond to dynamics at different energy scales associated
with the Hubbard bands and the quasiparticle peaks, respectively.
For simplicity, consider the case $N = 2$ at integer-filling $n = 1$
(the generalization for different $N$ and $[n]$ is straightforward).
At $T = 0$, the positive and negative frequency sides
of a correlator $A_{XY}$ directly correspond to the
Fourier transforms of $\la X(t) Y(0) \ra$ and $\la X^\dagger (t) Y^\dagger (0)\ra^*$,
respectively.
Hence, for example, the contribution to the UHB by $A_{dd^\dagger}$ corresponds to the dynamics
of a doublon excitation added at time $0$ and
then again removed at time $t$.
Conversely, the low-energy feature centered at $\omega = -\omega_s < 0$
means that a single spin remaining after
\emph{removing} a doublon at time~$0$ undergoes
a time evolution governed by the spin-like
collective mode with energy scale $\omega_s$,
until a doublon is regenerated on top of the spin at time $t$.
The features of $A_{h^\dagger h}$ can be explained
analogously by swapping the roles of doublon and holon.
On the other hand, the doublon-holon correlators $A_{dh} (\omega) = A_{h^\dagger d^\dagger} (\omega)$ mainly contribute to the quasiparticle peak, 
rather than to the Hubbard bands.
It means that the doublon and holon excitations are
combined at low energies to build quasiparticles.

Now we demonstrate that the asymmetry of the quasiparticle peak
in \Fig{fig:A} originates from striking differences between the
doublon and holon spectra, shown in \Fig{fig:DH}. These differences
stem from the strong asymmetry in energy cost for doublon and
holon excitations, due to large $U$, despite the low
level of particle doping $(n - [n])/ N < 0.03$.
In the slightly particle-doped regime, the UHB (LHB)
originating from local doublon (holon) excitation lies
close to (far from) the Fermi level.  Due to this strong asymmetry, the
metallic ground state in this particle-doped Mott insulator contains
much more doublons than holons;
correspondingly, $A_{dd^\dagger}(\omega = 0)$ is higher than
$A_{h^\dagger h}(\omega = 0)$ by more than an order
of magnitude. (This fact is evoked when constructing the $t$-$J$
model~\cite{Harris1967,*Chao1977,*MacDonald1988,*Eskes1994,*Eskes1994a},
a widely used effective low-energy model for a Mott insulator,
which in the case of hole-doping completely neglects
double-occupancy.) In any case, the metallic ground state
should contain delocalized doublon and/or holon fluctuations
since otherwise the system would be insulating.  Such fluctuations can
optimize the total energy, by decreasing kinetic energy more than
the related increase in Coulomb interaction energy.
Finally, the differences in  
spectral strength between $A_{dd^\dagger}$ and
$A_{h^\dagger h}$ in the quasiparticle peak regime,
combined with the fact that
the contributions of $A_{dd^\dagger}$
($A_{h^\dagger h}$) to the quasiparticle peak are
centered at $\omega = -\omega_s$ ($+\omega_s$),
necessarily cause the asymmetry of the
quasiparticle peak in $A (\omega)$ [cf.~\Eq{eq:Aomega}].

\section{Conclusion}
\label{sec:conclusion}

We have investigated the compressibility in the metallic and
paramagnetic insulating phases along the filling-driven Mott
transition of the $\mr{SU}(N)$ Hubbard model.  The compressibility
$\tilde{\kappa}$ vs.~the occupation number $n$ exhibits distinct
behaviours depending on temperature: (i) Below the critical
temperature $T_c$, $\tilde{\kappa}(n)$ discontinuously drops to zero
at integer $n$, as the manifestation of the Mott transition.  (ii)
Above $T_c$, the curve $\tilde{\kappa}(n)$ is continuous, since the
evolution between the metallic and insulating phases is now a
crossover, not a phase transition.  (iii) Near $T_c$, in the
metallic phase close to the Mott insulating phase, $\tilde{\kappa}$
shows a prominent enhancement, which directly coincides with
the thermal suppression of the quasiparticle peak.  The quasiparticle
peak represents the metallicity, in that it
hosts low-energy charge fluctuations and supports non-integer
occupation, while the absence of the quasiparticle peak leads to the
Mottness that allows only integer occupation.

We have also shown that,
in the vicinity of the filling-driven Mott transition,
the asymmetric position of the Hubbard bands and the asymmetric shape of the quasiparticle peak 
have the same origin:
different energy cost of doublon and holon excitations.

While we have focused on the paramagnetic phases in
this work, magnetic ordering such as anti-ferromagnetism
can occur in experiments,
as demonstrated in \Ref{Mazurenko2017}.
To describe anti-ferromagnetism,
it is necessary to go beyond the single-site setting
of DMFT which we employ here,
by using, e.g., bipartite lattice setting of DMFT~\cite{Chitra1999,Zitzler2004,Yanatori2016}
or real-space DMFT~\cite{Helmes2008}.
It would be interesting to study the compressibility
in the presence of magnetic ordering, yet 
this is beyond the scope of this work.

For the purpose of this paper,
the paramagnetic Mott transition may be achieved
in ultracold atom experiments by tuning
the critical temperature of the magnetic transition
significantly below the paramagnetic transition.
The critical temperature of the magnetic transition can be lowered
by having frustration in the system,
such as next-nearest-neighbour hopping or non-bipartite (e.g., triangular) lattice.
Another option is to increase the number $N$ of flavors,
which increases the critical temperature of the paramagnetic
transition~\cite{Bluemer2013,Yanatori2016} (see also
Figs.~\ref{fig:T0_n} and \ref{fig:Tdep_dn} in that the coexistence
region gets wider with larger $N$), yet decreases
the critical temperature for the magnetic transition
(e.g., according to \Ref{Yanatori2016},
the former becomes larger than the latter for $N \geq 6$).  This option
is appealing in that, for larger $N$, lower system temperatures are
accessible, since the Pomeranchuk effect, a mechanism to cool down
cold atoms, becomes stronger~\cite{Hazzard2012,Taie2012,Hofrichter2016}.

We thank S.~F\"olling, A.~Koga, and G.~Kotliar for fruitful discussions.
This work was supported by Nanosystems Initiative Munich.
S.L.~acknowledges support from the Alexander von Humboldt Foundation, the Carl Friedrich von Siemens Foundation,
and German-Israeli Foundation for Scientific Research and Development,
A.W.~from DFG WE4819/2-1.


\begin{thebibliography}{45}%
\makeatletter
\providecommand \@ifxundefined [1]{%
 \@ifx{#1\undefined}
}%
\providecommand \@ifnum [1]{%
 \ifnum #1\expandafter \@firstoftwo
 \else \expandafter \@secondoftwo
 \fi
}%
\providecommand \@ifx [1]{%
 \ifx #1\expandafter \@firstoftwo
 \else \expandafter \@secondoftwo
 \fi
}%
\providecommand \natexlab [1]{#1}%
\providecommand \enquote  [1]{``#1''}%
\providecommand \bibnamefont  [1]{#1}%
\providecommand \bibfnamefont [1]{#1}%
\providecommand \citenamefont [1]{#1}%
\providecommand \href@noop [0]{\@secondoftwo}%
\providecommand \href [0]{\begingroup \@sanitize@url \@href}%
\providecommand \@href[1]{\@@startlink{#1}\@@href}%
\providecommand \@@href[1]{\endgroup#1\@@endlink}%
\providecommand \@sanitize@url [0]{\catcode `\\12\catcode `\$12\catcode
  `\&12\catcode `\#12\catcode `\^12\catcode `\_12\catcode `\%12\relax}%
\providecommand \@@startlink[1]{}%
\providecommand \@@endlink[0]{}%
\providecommand \url  [0]{\begingroup\@sanitize@url \@url }%
\providecommand \@url [1]{\endgroup\@href {#1}{\urlprefix }}%
\providecommand \urlprefix  [0]{URL }%
\providecommand \Eprint [0]{\href }%
\providecommand \doibase [0]{http://dx.doi.org/}%
\providecommand \selectlanguage [0]{\@gobble}%
\providecommand \bibinfo  [0]{\@secondoftwo}%
\providecommand \bibfield  [0]{\@secondoftwo}%
\providecommand \translation [1]{[#1]}%
\providecommand \BibitemOpen [0]{}%
\providecommand \bibitemStop [0]{}%
\providecommand \bibitemNoStop [0]{.\EOS\space}%
\providecommand \EOS [0]{\spacefactor3000\relax}%
\providecommand \BibitemShut  [1]{\csname bibitem#1\endcsname}%
\let\auto@bib@innerbib\@empty
\bibitem [{\citenamefont {Imada}\ \emph {et~al.}(1998)\citenamefont {Imada},
  \citenamefont {Fujimori},\ and\ \citenamefont {Tokura}}]{Imada1998}%
  \BibitemOpen
  \bibfield  {author} {\bibinfo {author} {\bibfnamefont {M.}~\bibnamefont
  {Imada}}, \bibinfo {author} {\bibfnamefont {A.}~\bibnamefont {Fujimori}}, \
  and\ \bibinfo {author} {\bibfnamefont {Y.}~\bibnamefont {Tokura}},\ }\href
  {\doibase 10.1103/RevModPhys.70.1039} {\bibfield  {journal} {\bibinfo
  {journal} {Rev. Mod. Phys.}\ }\textbf {\bibinfo {volume} {70}},\ \bibinfo
  {pages} {1039} (\bibinfo {year} {1998})}\BibitemShut {NoStop}%
\bibitem [{\citenamefont {Fisher}\ \emph {et~al.}(1995)\citenamefont {Fisher},
  \citenamefont {Kotliar},\ and\ \citenamefont {Moeller}}]{Fisher1995}%
  \BibitemOpen
  \bibfield  {author} {\bibinfo {author} {\bibfnamefont {D.~S.}\ \bibnamefont
  {Fisher}}, \bibinfo {author} {\bibfnamefont {G.}~\bibnamefont {Kotliar}}, \
  and\ \bibinfo {author} {\bibfnamefont {G.}~\bibnamefont {Moeller}},\ }\href
  {\doibase 10.1103/PhysRevB.52.17112} {\bibfield  {journal} {\bibinfo
  {journal} {Phys. Rev. B}\ }\textbf {\bibinfo {volume} {52}},\ \bibinfo
  {pages} {17112} (\bibinfo {year} {1995})}\BibitemShut {NoStop}%
\bibitem [{\citenamefont {Kajueter}\ \emph {et~al.}(1996)\citenamefont
  {Kajueter}, \citenamefont {Kotliar},\ and\ \citenamefont
  {Moeller}}]{Kajueter1996}%
  \BibitemOpen
  \bibfield  {author} {\bibinfo {author} {\bibfnamefont {H.}~\bibnamefont
  {Kajueter}}, \bibinfo {author} {\bibfnamefont {G.}~\bibnamefont {Kotliar}}, \
  and\ \bibinfo {author} {\bibfnamefont {G.}~\bibnamefont {Moeller}},\ }\href
  {\doibase 10.1103/PhysRevB.53.16214} {\bibfield  {journal} {\bibinfo
  {journal} {Phys. Rev. B}\ }\textbf {\bibinfo {volume} {53}},\ \bibinfo
  {pages} {16214} (\bibinfo {year} {1996})}\BibitemShut {NoStop}%
\bibitem [{\citenamefont {Rozenberg}(1997)}]{Rozenberg1997}%
  \BibitemOpen
  \bibfield  {author} {\bibinfo {author} {\bibfnamefont {M.~J.}\ \bibnamefont
  {Rozenberg}},\ }\href {\doibase 10.1103/PhysRevB.55.R4855} {\bibfield
  {journal} {\bibinfo  {journal} {Phys. Rev. B}\ }\textbf {\bibinfo {volume}
  {55}},\ \bibinfo {pages} {R4855} (\bibinfo {year} {1997})}\BibitemShut
  {NoStop}%
\bibitem [{\citenamefont {Kotliar}\ \emph {et~al.}(2002)\citenamefont
  {Kotliar}, \citenamefont {Murthy},\ and\ \citenamefont
  {Rozenberg}}]{Kotliar2002}%
  \BibitemOpen
  \bibfield  {author} {\bibinfo {author} {\bibfnamefont {G.}~\bibnamefont
  {Kotliar}}, \bibinfo {author} {\bibfnamefont {S.}~\bibnamefont {Murthy}}, \
  and\ \bibinfo {author} {\bibfnamefont {M.~J.}\ \bibnamefont {Rozenberg}},\
  }\href {\doibase 10.1103/PhysRevLett.89.046401} {\bibfield  {journal}
  {\bibinfo  {journal} {Phys. Rev. Lett.}\ }\textbf {\bibinfo {volume} {89}},\
  \bibinfo {pages} {046401} (\bibinfo {year} {2002})}\BibitemShut {NoStop}%
\bibitem [{\citenamefont {Werner}\ and\ \citenamefont
  {Millis}(2007)}]{Werner2007}%
  \BibitemOpen
  \bibfield  {author} {\bibinfo {author} {\bibfnamefont {P.}~\bibnamefont
  {Werner}}\ and\ \bibinfo {author} {\bibfnamefont {A.~J.}\ \bibnamefont
  {Millis}},\ }\href {\doibase 10.1103/PhysRevB.75.085108} {\bibfield
  {journal} {\bibinfo  {journal} {Phys. Rev. B}\ }\textbf {\bibinfo {volume}
  {75}},\ \bibinfo {pages} {085108} (\bibinfo {year} {2007})}\BibitemShut
  {NoStop}%
\bibitem [{\citenamefont {Gorelik}\ and\ \citenamefont
  {Bl\"umer}(2009)}]{Gorelik2009}%
  \BibitemOpen
  \bibfield  {author} {\bibinfo {author} {\bibfnamefont {E.~V.}\ \bibnamefont
  {Gorelik}}\ and\ \bibinfo {author} {\bibfnamefont {N.}~\bibnamefont
  {Bl\"umer}},\ }\href {\doibase 10.1103/PhysRevA.80.051602} {\bibfield
  {journal} {\bibinfo  {journal} {Phys. Rev. A}\ }\textbf {\bibinfo {volume}
  {80}},\ \bibinfo {pages} {051602} (\bibinfo {year} {2009})}\BibitemShut
  {NoStop}%
\bibitem [{\citenamefont {J\"ordens}\ \emph {et~al.}(2008)\citenamefont
  {J\"ordens}, \citenamefont {Strohmaier}, \citenamefont {Gunter},
  \citenamefont {Moritz},\ and\ \citenamefont {Esslinger}}]{Jordens2008}%
  \BibitemOpen
  \bibfield  {author} {\bibinfo {author} {\bibfnamefont {R.}~\bibnamefont
  {J\"ordens}}, \bibinfo {author} {\bibfnamefont {N.}~\bibnamefont
  {Strohmaier}}, \bibinfo {author} {\bibfnamefont {K.}~\bibnamefont {Gunter}},
  \bibinfo {author} {\bibfnamefont {H.}~\bibnamefont {Moritz}}, \ and\ \bibinfo
  {author} {\bibfnamefont {T.}~\bibnamefont {Esslinger}},\ }\href
  {http://dx.doi.org/10.1038/nature07244} {\bibfield  {journal} {\bibinfo
  {journal} {Nature}\ }\textbf {\bibinfo {volume} {455}},\ \bibinfo {pages}
  {204} (\bibinfo {year} {2008})}\BibitemShut {NoStop}%
\bibitem [{\citenamefont {Schneider}\ \emph {et~al.}(2008)\citenamefont
  {Schneider}, \citenamefont {Hackerm{\"u}ller}, \citenamefont {Will},
  \citenamefont {Best}, \citenamefont {Bloch}, \citenamefont {Costi},
  \citenamefont {Helmes}, \citenamefont {Rasch},\ and\ \citenamefont
  {Rosch}}]{Schneider2008}%
  \BibitemOpen
  \bibfield  {author} {\bibinfo {author} {\bibfnamefont {U.}~\bibnamefont
  {Schneider}}, \bibinfo {author} {\bibfnamefont {L.}~\bibnamefont
  {Hackerm{\"u}ller}}, \bibinfo {author} {\bibfnamefont {S.}~\bibnamefont
  {Will}}, \bibinfo {author} {\bibfnamefont {T.}~\bibnamefont {Best}}, \bibinfo
  {author} {\bibfnamefont {I.}~\bibnamefont {Bloch}}, \bibinfo {author}
  {\bibfnamefont {T.~A.}\ \bibnamefont {Costi}}, \bibinfo {author}
  {\bibfnamefont {R.~W.}\ \bibnamefont {Helmes}}, \bibinfo {author}
  {\bibfnamefont {D.}~\bibnamefont {Rasch}}, \ and\ \bibinfo {author}
  {\bibfnamefont {A.}~\bibnamefont {Rosch}},\ }\href {\doibase
  10.1126/science.1165449} {\bibfield  {journal} {\bibinfo  {journal}
  {Science}\ }\textbf {\bibinfo {volume} {322}},\ \bibinfo {pages} {1520}
  (\bibinfo {year} {2008})}\BibitemShut {NoStop}%
\bibitem [{\citenamefont {Taie}\ \emph {et~al.}(2012)\citenamefont {Taie},
  \citenamefont {Yamazaki}, \citenamefont {Sugawa},\ and\ \citenamefont
  {Takahashi}}]{Taie2012}%
  \BibitemOpen
  \bibfield  {author} {\bibinfo {author} {\bibfnamefont {S.}~\bibnamefont
  {Taie}}, \bibinfo {author} {\bibfnamefont {R.}~\bibnamefont {Yamazaki}},
  \bibinfo {author} {\bibfnamefont {S.}~\bibnamefont {Sugawa}}, \ and\ \bibinfo
  {author} {\bibfnamefont {Y.}~\bibnamefont {Takahashi}},\ }\href
  {http://dx.doi.org/10.1038/nphys2430} {\bibfield  {journal} {\bibinfo
  {journal} {Nat. Phys.}\ }\textbf {\bibinfo {volume} {8}},\ \bibinfo {pages}
  {825} (\bibinfo {year} {2012})}\BibitemShut {NoStop}%
\bibitem [{\citenamefont {Duarte}\ \emph {et~al.}(2015)\citenamefont {Duarte},
  \citenamefont {Hart}, \citenamefont {Yang}, \citenamefont {Liu},
  \citenamefont {Paiva}, \citenamefont {Khatami}, \citenamefont {Scalettar},
  \citenamefont {Trivedi},\ and\ \citenamefont {Hulet}}]{Duarte2015}%
  \BibitemOpen
  \bibfield  {author} {\bibinfo {author} {\bibfnamefont {P.~M.}\ \bibnamefont
  {Duarte}}, \bibinfo {author} {\bibfnamefont {R.~A.}\ \bibnamefont {Hart}},
  \bibinfo {author} {\bibfnamefont {T.-L.}\ \bibnamefont {Yang}}, \bibinfo
  {author} {\bibfnamefont {X.}~\bibnamefont {Liu}}, \bibinfo {author}
  {\bibfnamefont {T.}~\bibnamefont {Paiva}}, \bibinfo {author} {\bibfnamefont
  {E.}~\bibnamefont {Khatami}}, \bibinfo {author} {\bibfnamefont {R.~T.}\
  \bibnamefont {Scalettar}}, \bibinfo {author} {\bibfnamefont {N.}~\bibnamefont
  {Trivedi}}, \ and\ \bibinfo {author} {\bibfnamefont {R.~G.}\ \bibnamefont
  {Hulet}},\ }\href {\doibase 10.1103/PhysRevLett.114.070403} {\bibfield
  {journal} {\bibinfo  {journal} {Phys. Rev. Lett.}\ }\textbf {\bibinfo
  {volume} {114}},\ \bibinfo {pages} {070403} (\bibinfo {year}
  {2015})}\BibitemShut {NoStop}%
\bibitem [{\citenamefont {Hofrichter}\ \emph {et~al.}(2016)\citenamefont
  {Hofrichter}, \citenamefont {Riegger}, \citenamefont {Scazza}, \citenamefont
  {H\"ofer}, \citenamefont {Fernandes}, \citenamefont {Bloch},\ and\
  \citenamefont {F\"olling}}]{Hofrichter2016}%
  \BibitemOpen
  \bibfield  {author} {\bibinfo {author} {\bibfnamefont {C.}~\bibnamefont
  {Hofrichter}}, \bibinfo {author} {\bibfnamefont {L.}~\bibnamefont {Riegger}},
  \bibinfo {author} {\bibfnamefont {F.}~\bibnamefont {Scazza}}, \bibinfo
  {author} {\bibfnamefont {M.}~\bibnamefont {H\"ofer}}, \bibinfo {author}
  {\bibfnamefont {D.~R.}\ \bibnamefont {Fernandes}}, \bibinfo {author}
  {\bibfnamefont {I.}~\bibnamefont {Bloch}}, \ and\ \bibinfo {author}
  {\bibfnamefont {S.}~\bibnamefont {F\"olling}},\ }\href {\doibase
  10.1103/PhysRevX.6.021030} {\bibfield  {journal} {\bibinfo  {journal} {Phys.
  Rev. X}\ }\textbf {\bibinfo {volume} {6}},\ \bibinfo {pages} {021030}
  (\bibinfo {year} {2016})}\BibitemShut {NoStop}%
\bibitem [{\citenamefont {Bl\"umer}(2002)}]{Bluemer2002}%
  \BibitemOpen
  \bibfield  {author} {\bibinfo {author} {\bibfnamefont {N.}~\bibnamefont
  {Bl\"umer}},\ }\href@noop {} {Ph.D. thesis},\ \bibinfo  {school}
  {Universit\"at Augsburg} (\bibinfo {year} {2002})\BibitemShut {NoStop}%
\bibitem [{\citenamefont {Bulla}\ \emph {et~al.}(2001)\citenamefont {Bulla},
  \citenamefont {Costi},\ and\ \citenamefont {Vollhardt}}]{Bulla2001}%
  \BibitemOpen
  \bibfield  {author} {\bibinfo {author} {\bibfnamefont {R.}~\bibnamefont
  {Bulla}}, \bibinfo {author} {\bibfnamefont {T.~A.}\ \bibnamefont {Costi}}, \
  and\ \bibinfo {author} {\bibfnamefont {D.}~\bibnamefont {Vollhardt}},\ }\href
  {\doibase 10.1103/PhysRevB.64.045103} {\bibfield  {journal} {\bibinfo
  {journal} {Phys. Rev. B}\ }\textbf {\bibinfo {volume} {64}},\ \bibinfo
  {pages} {045103} (\bibinfo {year} {2001})}\BibitemShut {NoStop}%
\bibitem [{\citenamefont {Yanatori}\ and\ \citenamefont
  {Koga}(2016)}]{Yanatori2016}%
  \BibitemOpen
  \bibfield  {author} {\bibinfo {author} {\bibfnamefont {H.}~\bibnamefont
  {Yanatori}}\ and\ \bibinfo {author} {\bibfnamefont {A.}~\bibnamefont
  {Koga}},\ }\href {\doibase 10.1103/PhysRevB.94.041110} {\bibfield  {journal}
  {\bibinfo  {journal} {Phys. Rev. B}\ }\textbf {\bibinfo {volume} {94}},\
  \bibinfo {pages} {041110} (\bibinfo {year} {2016})}\BibitemShut {NoStop}%
\bibitem [{\citenamefont {Mazurenko}\ \emph {et~al.}(2017)\citenamefont
  {Mazurenko}, \citenamefont {Chiu}, \citenamefont {Ji}, \citenamefont
  {Parsons}, \citenamefont {Kan{\'a}sz-Nagy}, \citenamefont {Schmidt},
  \citenamefont {Grusdt}, \citenamefont {Demler}, \citenamefont {Greif},\ and\
  \citenamefont {Greiner}}]{Mazurenko2017}%
  \BibitemOpen
  \bibfield  {author} {\bibinfo {author} {\bibfnamefont {A.}~\bibnamefont
  {Mazurenko}}, \bibinfo {author} {\bibfnamefont {C.~S.}\ \bibnamefont {Chiu}},
  \bibinfo {author} {\bibfnamefont {G.}~\bibnamefont {Ji}}, \bibinfo {author}
  {\bibfnamefont {M.~F.}\ \bibnamefont {Parsons}}, \bibinfo {author}
  {\bibfnamefont {M.}~\bibnamefont {Kan{\'a}sz-Nagy}}, \bibinfo {author}
  {\bibfnamefont {R.}~\bibnamefont {Schmidt}}, \bibinfo {author} {\bibfnamefont
  {F.}~\bibnamefont {Grusdt}}, \bibinfo {author} {\bibfnamefont
  {E.}~\bibnamefont {Demler}}, \bibinfo {author} {\bibfnamefont
  {D.}~\bibnamefont {Greif}}, \ and\ \bibinfo {author} {\bibfnamefont
  {M.}~\bibnamefont {Greiner}},\ }\href {http://dx.doi.org/10.1038/nature22362}
  {\bibfield  {journal} {\bibinfo  {journal} {Nature}\ }\textbf {\bibinfo
  {volume} {545}},\ \bibinfo {pages} {462} (\bibinfo {year}
  {2017})}\BibitemShut {NoStop}%
\bibitem [{\citenamefont {Georges}\ \emph {et~al.}(1996)\citenamefont
  {Georges}, \citenamefont {Kotliar}, \citenamefont {Krauth},\ and\
  \citenamefont {Rozenberg}}]{Georges1996}%
  \BibitemOpen
  \bibfield  {author} {\bibinfo {author} {\bibfnamefont {A.}~\bibnamefont
  {Georges}}, \bibinfo {author} {\bibfnamefont {G.}~\bibnamefont {Kotliar}},
  \bibinfo {author} {\bibfnamefont {W.}~\bibnamefont {Krauth}}, \ and\ \bibinfo
  {author} {\bibfnamefont {M.~J.}\ \bibnamefont {Rozenberg}},\ }\href {\doibase
  10.1103/RevModPhys.68.13} {\bibfield  {journal} {\bibinfo  {journal} {Rev.
  Mod. Phys.}\ }\textbf {\bibinfo {volume} {68}},\ \bibinfo {pages} {13}
  (\bibinfo {year} {1996})}\BibitemShut {NoStop}%
\bibitem [{\citenamefont {Kotliar}\ \emph {et~al.}(2006)\citenamefont
  {Kotliar}, \citenamefont {Savrasov}, \citenamefont {Haule}, \citenamefont
  {Oudovenko}, \citenamefont {Parcollet},\ and\ \citenamefont
  {Marianetti}}]{Kotliar2006}%
  \BibitemOpen
  \bibfield  {author} {\bibinfo {author} {\bibfnamefont {G.}~\bibnamefont
  {Kotliar}}, \bibinfo {author} {\bibfnamefont {S.~Y.}\ \bibnamefont
  {Savrasov}}, \bibinfo {author} {\bibfnamefont {K.}~\bibnamefont {Haule}},
  \bibinfo {author} {\bibfnamefont {V.~S.}\ \bibnamefont {Oudovenko}}, \bibinfo
  {author} {\bibfnamefont {O.}~\bibnamefont {Parcollet}}, \ and\ \bibinfo
  {author} {\bibfnamefont {C.~A.}\ \bibnamefont {Marianetti}},\ }\href
  {\doibase 10.1103/RevModPhys.78.865} {\bibfield  {journal} {\bibinfo
  {journal} {Rev. Mod. Phys.}\ }\textbf {\bibinfo {volume} {78}},\ \bibinfo
  {pages} {865} (\bibinfo {year} {2006})}\BibitemShut {NoStop}%
\bibitem [{\citenamefont {Wilson}(1975)}]{Wilson1975}%
  \BibitemOpen
  \bibfield  {author} {\bibinfo {author} {\bibfnamefont {K.~G.}\ \bibnamefont
  {Wilson}},\ }\href {\doibase 10.1103/RevModPhys.47.773} {\bibfield  {journal}
  {\bibinfo  {journal} {Rev. Mod. Phys.}\ }\textbf {\bibinfo {volume} {47}},\
  \bibinfo {pages} {773} (\bibinfo {year} {1975})}\BibitemShut {NoStop}%
\bibitem [{\citenamefont {Bulla}\ \emph {et~al.}(2008)\citenamefont {Bulla},
  \citenamefont {Costi},\ and\ \citenamefont {Pruschke}}]{Bulla2008}%
  \BibitemOpen
  \bibfield  {author} {\bibinfo {author} {\bibfnamefont {R.}~\bibnamefont
  {Bulla}}, \bibinfo {author} {\bibfnamefont {T.~A.}\ \bibnamefont {Costi}}, \
  and\ \bibinfo {author} {\bibfnamefont {T.}~\bibnamefont {Pruschke}},\ }\href
  {\doibase 10.1103/RevModPhys.80.395} {\bibfield  {journal} {\bibinfo
  {journal} {Rev. Mod. Phys.}\ }\textbf {\bibinfo {volume} {80}},\ \bibinfo
  {pages} {395} (\bibinfo {year} {2008})}\BibitemShut {NoStop}%
\bibitem [{\citenamefont {Lee}\ \emph {et~al.}(2017{\natexlab{a}})\citenamefont
  {Lee}, \citenamefont {von Delft},\ and\ \citenamefont
  {Weichselbaum}}]{Lee2017}%
  \BibitemOpen
  \bibfield  {author} {\bibinfo {author} {\bibfnamefont {S.-S.~B.}\
  \bibnamefont {Lee}}, \bibinfo {author} {\bibfnamefont {J.}~\bibnamefont {von
  Delft}}, \ and\ \bibinfo {author} {\bibfnamefont {A.}~\bibnamefont
  {Weichselbaum}},\ }\href@noop {} {\  (\bibinfo {year}
  {2017}{\natexlab{a}})},\ \bibinfo {note} {submitted to Phys. Rev. Lett.},\
  \Eprint {http://arxiv.org/abs/1705.03910} {arXiv:1705.03910} \BibitemShut
  {NoStop}%
\bibitem [{\citenamefont {Lee}\ \emph {et~al.}(2017{\natexlab{b}})\citenamefont
  {Lee}, \citenamefont {von Delft},\ and\ \citenamefont
  {Weichselbaum}}]{Lee2017a}%
  \BibitemOpen
  \bibfield  {author} {\bibinfo {author} {\bibfnamefont {S.-S.~B.}\
  \bibnamefont {Lee}}, \bibinfo {author} {\bibfnamefont {J.}~\bibnamefont {von
  Delft}}, \ and\ \bibinfo {author} {\bibfnamefont {A.}~\bibnamefont
  {Weichselbaum}},\ }\href@noop {} {\  (\bibinfo {year}
  {2017}{\natexlab{b}})},\ \bibinfo {note} {submitted to Phys. Rev. B},\
  \Eprint {http://arxiv.org/abs/1710.01171} {arXiv:1710.01171} \BibitemShut
  {NoStop}%
\bibitem [{\citenamefont {Helmes}\ \emph {et~al.}(2008)\citenamefont {Helmes},
  \citenamefont {Costi},\ and\ \citenamefont {Rosch}}]{Helmes2008}%
  \BibitemOpen
  \bibfield  {author} {\bibinfo {author} {\bibfnamefont {R.~W.}\ \bibnamefont
  {Helmes}}, \bibinfo {author} {\bibfnamefont {T.~A.}\ \bibnamefont {Costi}}, \
  and\ \bibinfo {author} {\bibfnamefont {A.}~\bibnamefont {Rosch}},\ }\href
  {\doibase 10.1103/PhysRevLett.100.056403} {\bibfield  {journal} {\bibinfo
  {journal} {Phys. Rev. Lett.}\ }\textbf {\bibinfo {volume} {100}},\ \bibinfo
  {pages} {056403} (\bibinfo {year} {2008})}\BibitemShut {NoStop}%
\bibitem [{\citenamefont {Metzner}\ and\ \citenamefont
  {Vollhardt}(1989)}]{Metzner1989}%
  \BibitemOpen
  \bibfield  {author} {\bibinfo {author} {\bibfnamefont {W.}~\bibnamefont
  {Metzner}}\ and\ \bibinfo {author} {\bibfnamefont {D.}~\bibnamefont
  {Vollhardt}},\ }\href {\doibase 10.1103/PhysRevLett.62.324} {\bibfield
  {journal} {\bibinfo  {journal} {Phys. Rev. Lett.}\ }\textbf {\bibinfo
  {volume} {62}},\ \bibinfo {pages} {324} (\bibinfo {year} {1989})}\BibitemShut
  {NoStop}%
\bibitem [{\citenamefont {Weichselbaum}\ and\ \citenamefont {von
  Delft}(2007)}]{Weichselbaum2007}%
  \BibitemOpen
  \bibfield  {author} {\bibinfo {author} {\bibfnamefont {A.}~\bibnamefont
  {Weichselbaum}}\ and\ \bibinfo {author} {\bibfnamefont {J.}~\bibnamefont {von
  Delft}},\ }\href {\doibase 10.1103/PhysRevLett.99.076402} {\bibfield
  {journal} {\bibinfo  {journal} {Phys. Rev. Lett.}\ }\textbf {\bibinfo
  {volume} {99}},\ \bibinfo {pages} {076402} (\bibinfo {year}
  {2007})}\BibitemShut {NoStop}%
\bibitem [{\citenamefont
  {Weichselbaum}(2012{\natexlab{a}})}]{Weichselbaum2012:mps}%
  \BibitemOpen
  \bibfield  {author} {\bibinfo {author} {\bibfnamefont {A.}~\bibnamefont
  {Weichselbaum}},\ }\href {\doibase 10.1103/PhysRevB.86.245124} {\bibfield
  {journal} {\bibinfo  {journal} {Phys. Rev. B}\ }\textbf {\bibinfo {volume}
  {86}},\ \bibinfo {pages} {245124} (\bibinfo {year}
  {2012}{\natexlab{a}})}\BibitemShut {NoStop}%
\bibitem [{\citenamefont {Anders}\ and\ \citenamefont
  {Schiller}(2005)}]{Anders2005}%
  \BibitemOpen
  \bibfield  {author} {\bibinfo {author} {\bibfnamefont {F.~B.}\ \bibnamefont
  {Anders}}\ and\ \bibinfo {author} {\bibfnamefont {A.}~\bibnamefont
  {Schiller}},\ }\href {\doibase 10.1103/PhysRevLett.95.196801} {\bibfield
  {journal} {\bibinfo  {journal} {Phys. Rev. Lett.}\ }\textbf {\bibinfo
  {volume} {95}},\ \bibinfo {pages} {196801} (\bibinfo {year}
  {2005})}\BibitemShut {NoStop}%
\bibitem [{\citenamefont {Anders}\ and\ \citenamefont
  {Schiller}(2006)}]{Anders2006}%
  \BibitemOpen
  \bibfield  {author} {\bibinfo {author} {\bibfnamefont {F.~B.}\ \bibnamefont
  {Anders}}\ and\ \bibinfo {author} {\bibfnamefont {A.}~\bibnamefont
  {Schiller}},\ }\href {\doibase 10.1103/PhysRevB.74.245113} {\bibfield
  {journal} {\bibinfo  {journal} {Phys. Rev. B}\ }\textbf {\bibinfo {volume}
  {74}},\ \bibinfo {pages} {245113} (\bibinfo {year} {2006})}\BibitemShut
  {NoStop}%
\bibitem [{\citenamefont {Lee}\ and\ \citenamefont
  {Weichselbaum}(2016)}]{Lee2016}%
  \BibitemOpen
  \bibfield  {author} {\bibinfo {author} {\bibfnamefont {S.-S.~B.}\
  \bibnamefont {Lee}}\ and\ \bibinfo {author} {\bibfnamefont {A.}~\bibnamefont
  {Weichselbaum}},\ }\href {\doibase 10.1103/PhysRevB.94.235127} {\bibfield
  {journal} {\bibinfo  {journal} {Phys. Rev. B}\ }\textbf {\bibinfo {volume}
  {94}},\ \bibinfo {pages} {235127} (\bibinfo {year} {2016})}\BibitemShut
  {NoStop}%
\bibitem [{\citenamefont
  {Weichselbaum}(2012{\natexlab{b}})}]{Weichselbaum2012:sym}%
  \BibitemOpen
  \bibfield  {author} {\bibinfo {author} {\bibfnamefont {A.}~\bibnamefont
  {Weichselbaum}},\ }\href {\doibase 10.1016/j.aop.2012.07.009} {\bibfield
  {journal} {\bibinfo  {journal} {Ann. Phys.}\ }\textbf {\bibinfo {volume}
  {327}},\ \bibinfo {pages} {2972} (\bibinfo {year}
  {2012}{\natexlab{b}})}\BibitemShut {NoStop}%
\bibitem [{\citenamefont {\v{Z}itko}\ and\ \citenamefont
  {Pruschke}(2009)}]{Zitko2009}%
  \BibitemOpen
  \bibfield  {author} {\bibinfo {author} {\bibfnamefont {R.}~\bibnamefont
  {\v{Z}itko}}\ and\ \bibinfo {author} {\bibfnamefont {T.}~\bibnamefont
  {Pruschke}},\ }\href {\doibase 10.1103/PhysRevB.79.085106} {\bibfield
  {journal} {\bibinfo  {journal} {Phys. Rev. B}\ }\textbf {\bibinfo {volume}
  {79}},\ \bibinfo {pages} {085106} (\bibinfo {year} {2009})}\BibitemShut
  {NoStop}%
\bibitem [{Note1()}]{Note1}%
  \BibitemOpen
  \bibinfo {note} {In the NRG, the convolution relation $n = N \DOTSI \intop
  \ilimits@ _{-\infty }^\infty \protect \ensuremath {\protect \mathrm
  {d}}\omega $ $A(\omega ) / (e^{\omega /T} + 1)$ holds when the local spectral
  function $A (\omega )$ is the discrete data in the Lehmann representation
  before broadening, not the continuous curve as in Fig.~\ref {fig:A}. Since
  the linear broadening~\cite {Weichselbaum2007,Lee2016} smooths out $A(\omega
  )$ for $|\omega | \lesssim T$, using the broadened $A(\omega )$ can introduce
  an artifact to the convolution relation.}\BibitemShut {Stop}%
\bibitem [{\citenamefont {Bulla}(1999)}]{Bulla1999}%
  \BibitemOpen
  \bibfield  {author} {\bibinfo {author} {\bibfnamefont {R.}~\bibnamefont
  {Bulla}},\ }\href {\doibase 10.1103/PhysRevLett.83.136} {\bibfield  {journal}
  {\bibinfo  {journal} {Phys. Rev. Lett.}\ }\textbf {\bibinfo {volume} {83}},\
  \bibinfo {pages} {136} (\bibinfo {year} {1999})}\BibitemShut {NoStop}%
\bibitem [{\citenamefont {\ifmmode~\bar{O}\else \={O}\fi{}no}\ \emph
  {et~al.}(2003)\citenamefont {\ifmmode~\bar{O}\else \={O}\fi{}no},
  \citenamefont {Potthoff},\ and\ \citenamefont {Bulla}}]{Ono2003}%
  \BibitemOpen
  \bibfield  {author} {\bibinfo {author} {\bibfnamefont {Y.}~\bibnamefont
  {\ifmmode~\bar{O}\else \={O}\fi{}no}}, \bibinfo {author} {\bibfnamefont
  {M.}~\bibnamefont {Potthoff}}, \ and\ \bibinfo {author} {\bibfnamefont
  {R.}~\bibnamefont {Bulla}},\ }\href {\doibase 10.1103/PhysRevB.67.035119}
  {\bibfield  {journal} {\bibinfo  {journal} {Phys. Rev. B}\ }\textbf {\bibinfo
  {volume} {67}},\ \bibinfo {pages} {035119} (\bibinfo {year}
  {2003})}\BibitemShut {NoStop}%
\bibitem [{\citenamefont {Inaba}\ \emph {et~al.}(2005)\citenamefont {Inaba},
  \citenamefont {Koga}, \citenamefont {Suga},\ and\ \citenamefont
  {Kawakami}}]{Inaba2005}%
  \BibitemOpen
  \bibfield  {author} {\bibinfo {author} {\bibfnamefont {K.}~\bibnamefont
  {Inaba}}, \bibinfo {author} {\bibfnamefont {A.}~\bibnamefont {Koga}},
  \bibinfo {author} {\bibfnamefont {S.-i.}\ \bibnamefont {Suga}}, \ and\
  \bibinfo {author} {\bibfnamefont {N.}~\bibnamefont {Kawakami}},\ }\href
  {\doibase 10.1103/PhysRevB.72.085112} {\bibfield  {journal} {\bibinfo
  {journal} {Phys. Rev. B}\ }\textbf {\bibinfo {volume} {72}},\ \bibinfo
  {pages} {085112} (\bibinfo {year} {2005})}\BibitemShut {NoStop}%
\bibitem [{\citenamefont {Bl\"umer}\ and\ \citenamefont
  {Gorelik}(2013)}]{Bluemer2013}%
  \BibitemOpen
  \bibfield  {author} {\bibinfo {author} {\bibfnamefont {N.}~\bibnamefont
  {Bl\"umer}}\ and\ \bibinfo {author} {\bibfnamefont {E.~V.}\ \bibnamefont
  {Gorelik}},\ }\href {\doibase 10.1103/PhysRevB.87.085115} {\bibfield
  {journal} {\bibinfo  {journal} {Phys. Rev. B}\ }\textbf {\bibinfo {volume}
  {87}},\ \bibinfo {pages} {085115} (\bibinfo {year} {2013})}\BibitemShut
  {NoStop}%
\bibitem [{\citenamefont {M{\"u}ller-Hartmann}(1989)}]{Mueller-Hartmann1989}%
  \BibitemOpen
  \bibfield  {author} {\bibinfo {author} {\bibfnamefont {E.}~\bibnamefont
  {M{\"u}ller-Hartmann}},\ }\href {\doibase 10.1007/BF01312686} {\bibfield
  {journal} {\bibinfo  {journal} {Z. Phys. B}\ }\textbf {\bibinfo {volume}
  {76}},\ \bibinfo {pages} {211} (\bibinfo {year} {1989})}\BibitemShut
  {NoStop}%
\bibitem [{\citenamefont {Harris}\ and\ \citenamefont
  {Lange}(1967)}]{Harris1967}%
  \BibitemOpen
  \bibfield  {author} {\bibinfo {author} {\bibfnamefont {A.~B.}\ \bibnamefont
  {Harris}}\ and\ \bibinfo {author} {\bibfnamefont {R.~V.}\ \bibnamefont
  {Lange}},\ }\href {\doibase 10.1103/PhysRev.157.295} {\bibfield  {journal}
  {\bibinfo  {journal} {Phys. Rev.}\ }\textbf {\bibinfo {volume} {157}},\
  \bibinfo {pages} {295} (\bibinfo {year} {1967})}\BibitemShut {NoStop}%
\bibitem [{\citenamefont {Chao}\ \emph {et~al.}(1977)\citenamefont {Chao},
  \citenamefont {Spalek},\ and\ \citenamefont {Oles}}]{Chao1977}%
  \BibitemOpen
  \bibfield  {author} {\bibinfo {author} {\bibfnamefont {K.~A.}\ \bibnamefont
  {Chao}}, \bibinfo {author} {\bibfnamefont {J.}~\bibnamefont {Spalek}}, \ and\
  \bibinfo {author} {\bibfnamefont {A.~M.}\ \bibnamefont {Oles}},\ }\href
  {http://stacks.iop.org/0022-3719/10/i=10/a=002} {\bibfield  {journal}
  {\bibinfo  {journal} {J. Phys. C}\ }\textbf {\bibinfo {volume} {10}},\
  \bibinfo {pages} {L271} (\bibinfo {year} {1977})}\BibitemShut {NoStop}%
\bibitem [{\citenamefont {MacDonald}\ \emph {et~al.}(1988)\citenamefont
  {MacDonald}, \citenamefont {Girvin},\ and\ \citenamefont
  {Yoshioka}}]{MacDonald1988}%
  \BibitemOpen
  \bibfield  {author} {\bibinfo {author} {\bibfnamefont {A.~H.}\ \bibnamefont
  {MacDonald}}, \bibinfo {author} {\bibfnamefont {S.~M.}\ \bibnamefont
  {Girvin}}, \ and\ \bibinfo {author} {\bibfnamefont {D.}~\bibnamefont
  {Yoshioka}},\ }\href {\doibase 10.1103/PhysRevB.37.9753} {\bibfield
  {journal} {\bibinfo  {journal} {Phys. Rev. B}\ }\textbf {\bibinfo {volume}
  {37}},\ \bibinfo {pages} {9753} (\bibinfo {year} {1988})}\BibitemShut
  {NoStop}%
\bibitem [{\citenamefont {Eskes}\ and\ \citenamefont
  {Ole\ifmmode~\acute{s}\else \'{s}\fi{}}(1994)}]{Eskes1994}%
  \BibitemOpen
  \bibfield  {author} {\bibinfo {author} {\bibfnamefont {H.}~\bibnamefont
  {Eskes}}\ and\ \bibinfo {author} {\bibfnamefont {A.~M.}\ \bibnamefont
  {Ole\ifmmode~\acute{s}\else \'{s}\fi{}}},\ }\href {\doibase
  10.1103/PhysRevLett.73.1279} {\bibfield  {journal} {\bibinfo  {journal}
  {Phys. Rev. Lett.}\ }\textbf {\bibinfo {volume} {73}},\ \bibinfo {pages}
  {1279} (\bibinfo {year} {1994})}\BibitemShut {NoStop}%
\bibitem [{\citenamefont {Eskes}\ \emph {et~al.}(1994)\citenamefont {Eskes},
  \citenamefont {Ole\ifmmode~\acute{s}\else \'{s}\fi{}}, \citenamefont
  {Meinders},\ and\ \citenamefont {Stephan}}]{Eskes1994a}%
  \BibitemOpen
  \bibfield  {author} {\bibinfo {author} {\bibfnamefont {H.}~\bibnamefont
  {Eskes}}, \bibinfo {author} {\bibfnamefont {A.~M.}\ \bibnamefont
  {Ole\ifmmode~\acute{s}\else \'{s}\fi{}}}, \bibinfo {author} {\bibfnamefont
  {M.~B.~J.}\ \bibnamefont {Meinders}}, \ and\ \bibinfo {author} {\bibfnamefont
  {W.}~\bibnamefont {Stephan}},\ }\href {\doibase 10.1103/PhysRevB.50.17980}
  {\bibfield  {journal} {\bibinfo  {journal} {Phys. Rev. B}\ }\textbf {\bibinfo
  {volume} {50}},\ \bibinfo {pages} {17980} (\bibinfo {year}
  {1994})}\BibitemShut {NoStop}%
\bibitem [{\citenamefont {Chitra}\ and\ \citenamefont
  {Kotliar}(1999)}]{Chitra1999}%
  \BibitemOpen
  \bibfield  {author} {\bibinfo {author} {\bibfnamefont {R.}~\bibnamefont
  {Chitra}}\ and\ \bibinfo {author} {\bibfnamefont {G.}~\bibnamefont
  {Kotliar}},\ }\href {\doibase 10.1103/PhysRevLett.83.2386} {\bibfield
  {journal} {\bibinfo  {journal} {Phys. Rev. Lett.}\ }\textbf {\bibinfo
  {volume} {83}},\ \bibinfo {pages} {2386} (\bibinfo {year}
  {1999})}\BibitemShut {NoStop}%
\bibitem [{\citenamefont {Zitzler}\ \emph {et~al.}(2004)\citenamefont
  {Zitzler}, \citenamefont {Tong}, \citenamefont {Pruschke},\ and\
  \citenamefont {Bulla}}]{Zitzler2004}%
  \BibitemOpen
  \bibfield  {author} {\bibinfo {author} {\bibfnamefont {R.}~\bibnamefont
  {Zitzler}}, \bibinfo {author} {\bibfnamefont {N.-H.}\ \bibnamefont {Tong}},
  \bibinfo {author} {\bibfnamefont {T.}~\bibnamefont {Pruschke}}, \ and\
  \bibinfo {author} {\bibfnamefont {R.}~\bibnamefont {Bulla}},\ }\href
  {\doibase 10.1103/PhysRevLett.93.016406} {\bibfield  {journal} {\bibinfo
  {journal} {Phys. Rev. Lett.}\ }\textbf {\bibinfo {volume} {93}},\ \bibinfo
  {pages} {016406} (\bibinfo {year} {2004})}\BibitemShut {NoStop}%
\bibitem [{\citenamefont {Hazzard}\ \emph {et~al.}(2012)\citenamefont
  {Hazzard}, \citenamefont {Gurarie}, \citenamefont {Hermele},\ and\
  \citenamefont {Rey}}]{Hazzard2012}%
  \BibitemOpen
  \bibfield  {author} {\bibinfo {author} {\bibfnamefont {K.~R.~A.}\
  \bibnamefont {Hazzard}}, \bibinfo {author} {\bibfnamefont {V.}~\bibnamefont
  {Gurarie}}, \bibinfo {author} {\bibfnamefont {M.}~\bibnamefont {Hermele}}, \
  and\ \bibinfo {author} {\bibfnamefont {A.~M.}\ \bibnamefont {Rey}},\ }\href
  {\doibase 10.1103/PhysRevA.85.041604} {\bibfield  {journal} {\bibinfo
  {journal} {Phys. Rev. A}\ }\textbf {\bibinfo {volume} {85}},\ \bibinfo
  {pages} {041604} (\bibinfo {year} {2012})}\BibitemShut {NoStop}%
\end{thebibliography}
 \newcommand{\noop}[1]{}

\end{document}